\documentclass[aps,nofootinbib,preprintnumbers,prd,onecolumn,12pt]{article}

\usepackage{amsmath}
\usepackage{subfig}
\usepackage{graphicx}
\usepackage{graphicx}
\usepackage[dvinames]{xcolor}
\usepackage{xspace}
\usepackage{url}

\newcommand{\Eqref}[1]{Eq.~\ref{#1}\xspace}

\def\Pmu         {\ensuremath{\upmu}\xspace}

\def\Ptau        {\ensuremath{\uptau}\xspace}

\def\Pmu         {\ensuremath{\mu}\xspace}

\def\Ptau        {\ensuremath{\tau}\xspace}

\def\tmmm        {\tau \to \mu \mu \mu}

\begin{document}
\begin{titlepage}

\begin{flushright}
{ \bf IFJ-PAN-IV-2016-24 }
\end{flushright}

\vspace{0.1cm}
\begin{center}
{\Huge \bf TAUOLA of $\Ptau$ lepton decays-- framework for hadronic currents, 
matrix elements and anomalous decays.}
\end{center}

\vspace*{1mm}

\begin{center}
{\bf M. Chrzaszcz$^{a,b}$, T. Przedzi\'nski$^{c}$, Z. Was$^{b}$ and J. Zaremba$^{b}$.}\\
{\em $^a$ Physik-Institute, University of Z\"urich, Winterthurerstrasse 190, CH-8057 Z\"urich, Switzerland}\\
{\em $^b$ Institute of Nuclear Physics Polish Academy of Sciences , PL-31342 Krakow, Poland}\\
{\em $^c$ Institute of Physics, Jagellonian University, Lojasiewicza 11, PL-30-348 Krakow, Poland}
\end{center}
\vspace{.1 cm}
\begin{center}
{\bf   ABSTRACT  }
\end{center}

We present an update of the Monte Carlo event generator {\tt TAUOLA}
for $\tau$ lepton decays, with substantially increased list of decay channels
and new initialization options.
The core of the program remains written in {\tt FORTRAN}
but necessary arrangements have been made to allow handling 
of the user-provided hadronic currents and matrix elements at 
the execution time. 
Such solution may simplify preparation of new hadronic currents 
and may be useful for fitting to the experimental data as well. 

We have implemented as default for {\tt TAUOLA} a set of hadronic currents,
which is compatible with the default initialization
used by BaBar collaboration. 
Options for currents available in previous releases are still stored 
in the code, sometimes left defunct or activated by internal flags only. 
The new version of the program, includes also implementation 
of Lepton Flavour Violating $\Ptau$ decays.

Finally, we present, as an example, 
 a set of {\tt C++} methods for handling user-provided 
currents, matrix elements or complete new decay channels initialization
which can be performed  at the program execution time.

\vfill %
\vspace{0.1 cm}
\begin{flushleft}
{   IFJ-PAN-IV-2016-24 \\
 Updated version as of May 2017}
\end{flushleft}

\vspace*{1mm}
\bigskip
\end{titlepage}

\section{Introduction}

The precision of present day $\tau$ leptons measurements is far better than
what theoretical models can offer.  We have discussed numerical consequences of this point 
in Ref.~\cite{Was:2015laa}.
The design of  Monte Carlo  generators simulating $\tau$ lepton decays, have to be experiment oriented:
the models describing hadronic effects of decays, are expected to be modified frequently.
The experiment interfaces parts of the code implementing
basic  physics principles such as of symmetry and finally
structures for data handling in simulation programs and in fitting should remain re-usable, 
to save time for the users and for easier debugging.
Some of the functionalities needed for the aforementioned purposes were already present in {\tt TAUOLA} but 
required modification to the code and recompilation. Now, we have adopted program structure, that the 
introduction of user-friendly methods to swap or add currents via pointers to user-defined functions is possible.
Information needed to handle these functions is coded  in {\tt FORTRAN common blocks}%
\footnote{All {\tt FORTRAN common blocks} can be accessed from {\tt C++} as
{\tt structs}. As such, we will use terms {\tt FORTRAN common blocks} and
{\tt C++ structs} interchangeably, depending on the context.}.
This is a good step toward initialization based on the lists of decay channels, which was started already 
many years ago \cite{Was:2000st} for the benefit of BaBar collaboration.

The $\tau$ leptons are the heaviest of all observed leptons. Their decays into hadrons constitute large fraction of all decay channels.
For their predictions results from intermediate energy scale QCD are needed.
 For such a regime theoretical 
predictions are of limited precision only. We will return to this point later, 
and let us recall our experience with work on Refs.~\cite{Shekhovtsova:2012ra,Nugent:2013hxa} only.
Also because of their large mass, $\tau$ leptons may potentially have large Yukawa couplings to 
New Particles or fields, and therefore be particularly sensitive to effects beyond 
the Standard Model (SM).
One of the cleanest theoretical way to discover New Physics
is via so called Lepton Flavour Violation processes.
These phenomena were already observed in the neutral lepton sector
via neutrino oscillations~\cite{Fukuda:1998mi,Ahmad:2002jz}.
In the charged lepton sector no such processes have been observed so far,
despite a large experimental effort~\cite{Amhis:2014hma}.
Also, sterile neutrinos seem to escape detection~\cite{Fukuda:2000np, Aaij:2014aba}.

The violation of charged lepton flavour (CLFV) is predicted in many extensions of the
SM (generically named as BSM theories).
The inclusion of the CLFV is usually straightforward and follows directly from the model's assumptions.
Among these theories are: supersymmetry (SUSY)~\cite{Aitchison:2005cf},
seesaw models~\cite{Melfo:2011nx},
little Higgs scenarios~\cite{Schmaltz:2002wx} and models with
four generations of fermions~\cite{braus4th_gen,Carpenter:2010bs}.
That is also why easy introduction of anomalous $\tau$ decay modes
becomes an important step of the program development.
Our paper describes how  such functionality can be achieved and 
ported for  both: {\tt FORTRAN} environments such as KKMC \cite{Jadach:1999vf}
and for {\tt C++} environments like {\tt Tauola Universal Interface}~\cite{Davidson:2010rw}.
Finally, archivization of $\tau$-decay initialization compatible with the one
used by BaBar collaboration was a  motivation for the 
 presented work.

Our paper is organized as follows. In Section \ref{sec:assumpt}
we recall basic assumptions enabling construction of the program.
Necessary approximations are presented. Section \ref{sec:data} documents
data structures on which architecture of the whole software relies on.
Methods essential for the organization of the program are documented there.
Rather minor changes
with respect to versions of Refs.~\cite{Was:2000st,Was:2004dg} 
were necessary. Program remains essentially in {\tt FORTRAN} only.
Section~\ref{sec:initializations} is devoted  to 
the new default {\tt TAUOLA} initialization for hadronic currents.
Minor comments on the possible variants are nonetheless given.
Section~\ref{sec:Exampleinst} describes a physics input for the matrix elements of the non-standard-model type.
An example of such anomalous $\tau$ decay mode is presented. Thanks to the new software organization,
such decay mode can be now easily introduced into {\tt TAUOLA} by the user.
Summary, Section~\ref{sec:summary}, closes the paper.

Appendices cover technical information related to the source code
described in this paper.
Appendix~\ref{app:interface_example} documents an example implementation
of the interface for adding new decay channels which is provided with the distribution source code.
Despite being a prototype all necessary features we expect the user may want are implemented. 
Appendix~\ref{app:install} provides instruction how present update
of {\tt TAUOLA} can be installed as well as examples of the program use. Furthermore, all steps required for installation into {\tt C++} environment
of {\tt Tauola Universal Interface} and {\tt TauSpinner} \cite{Davidson:2010rw} or into KKMC \cite{Jadach:1999vf}
environment, which remains predominantly in {\tt F77}, are given. 
Lastly, Appendix~\ref{app:prints} presents
the printout of information for all  decay channels provided with the distribution. 
This includes the branching ratios and flavours of stable decay products  
for the channels of the default program initialization. 
This includes also information for non initialized 
channels: placeholder channels available to be substituted by the user.

\section{Physics assumption}
\label{sec:assumpt}

Event generation in {\tt TAUOLA} is built out of several steps that involve
different types of theory calculations and thus provide different accuracy of the result.
In the leptonic decays all calculations are based on QED and
Fermi limit of electroweak interaction, which as we will see, is 
in general case, sufficiently precise.
Therefore, in the following description, we will concentrate only on semileptonic decays.
The event generation starts with a phase-space parametrization which
is exact and depends on masses of final state particles.
Formula~\ref{eq:ps} used directly is inefficient for a Monte
Carlo algorithm if sharp peaks are present due to resonances in the intermediate states.
The changes of variables affect the program efficiency
but the actual density of the phase space remains intact.
In this Section we concentrate on $\tau$ decays to three scalars and neutrino,
but principles described here are true for all other decay modes as well.
Phase space, $d{\rm Lips}$ (in the following, notation taken from Ref.~\cite{Jadach:1993hs} is used)

\begin{multline}
d{\rm Lips}(P;q_{1},q_{2},q_{3},q_{4})=
\frac{1}{2^{17}\pi^{8}}~\int^{Q^{2}_{max}}_{Q^{2}_{min}}dQ^{2}~
\int^{M^{2}_{2,max}}_{M^{2}_{2,min}}dM_{2}^{2}\\ \times \int
d\Omega_{4}\frac{\sqrt{\lambda(M_{2}^{2},Q^{2},m_{4}^{2})}}{M_{2}^{2}} \int
d\Omega_{3}\frac{\sqrt{\lambda(Q^{2},m_{3}^{2},M_{2}^{2})}}{Q^{2}} \int
d\Omega_{2}\frac{\sqrt{\lambda(M_{2}^{2},m_{2}^{2},m_{1}^{2})}}{M_{2}^{2}}\\
Q^{2}=(q_{1}+q_{2}+q_{3})^{2},~~~~~~M_{2}^{2}=(q_{1}+q_{2})^{2},\\
Q_{min}=m_{1}+m_{2}+m_{3}, ~~~~~~~Q_{max}=M-m_{4}\\
M_{2,min}=m_{1}+m_{2}, ~~~~~~M_{2,max}=Q-m_{3}\\
\label{eq:ps}
\end{multline}

is calculated independently of the matrix elements~\Eqref{eq:ME}.
Matrix element  consists of weak and hadronic currents.
Weak current is calculated for Fermi-like interaction,

\begin{equation}
\tau(P,s)\rightarrow\nu_{\tau}(N)X, ~~~~~~~~~~
{\cal M}=\frac{G}{\sqrt{2}}\bar{u}(N)\gamma^{\mu}(v+a\gamma_{5})u(P) \cdot J_{\mu}.
\label{eq:ME}
\end{equation}

Hadronic current $J_{\mu}$ depends on the momenta of all hadrons.
Such separation is valid for precision of about $\alpha/\pi \approx 0.2 \%$.
After straightforward calculation we obtain:

\begin{multline}
|{\cal M}|^{2}= G^{2}\frac{v^{2}+a^{2}}{2}( \omega + H_{\mu}s^{\mu} ),\\
\omega=P^{\mu}(\Pi_{\mu}-\gamma_{va}\Pi_{\mu}^{5}),\\
H_{\mu}=\frac{1}{M}(M^{2}\delta^{\nu}_{\mu}-P_{\mu}P^{\nu})(\Pi_{\nu}^{5}-\gamma_{va}\Pi_{\nu}),\\
\Pi_{\mu}=2[(J^{*}\cdot N)J_{\mu}+(J\cdot N)J_{\mu}^{*}-(J^{*}\cdot J)N_{\mu}],\\
\Pi^{5\mu}=2~ {\rm Im} ~\epsilon^{\mu\nu\rho\sigma}J^{*}_{\nu}J_{\rho}N_{\sigma},\\
\gamma_{va}=-\frac{2va}{v^{2}+a^{2}}\\
\label{eq:ME2}
\end{multline}


If a more general coupling
$v+a\gamma_{5}$ for the $\tau$ current and $\nu_{\tau}$ mass $m_{\nu} \neq 0$
are expected to be used in \Eqref{eq:ME2}, the following terms have to be added to $\omega$ and $H_{\mu}$:

\begin{multline}
\hat{\omega}=2\frac{v^{2}-a^{2}}{v^{2}+a^{2}}m_{\nu}M(J^{*} \cdot J), ~~~~
\hat{H}^{\mu}=-2\frac{v^{2}-a^{2}}{v^{2}+a^{2}}m_{\nu}~ {\rm Im}~\epsilon^{\mu\nu\rho\sigma}J_{\nu}^{*}J_{\rho}P_{\sigma}
\end{multline}


Nothing has to be changed with respect to ref.~\cite{Jadach:1993hs}.

In case of semileptonic decays hadronic current
is of the particular interest for the user, as it is the only source 
of substantial systematic errors.
There is no strict way of calculating it,
therefore user might be interested in obtaining them from fitting models to the data.
In such application a user-friendly interface is crucial.
A typical precision of the models is of order
of $1/N_c \approx 30\%$  or $1/N_c^2 \approx 10\%$ while experimental data precision is better than $0.1\%$ in most of the cases.

The differential partial width~\Eqref{eq:diff} for the semileptonic channels reads as
product of the phase-space, matrix element squared and flux factor.

\begin{equation}
d\Gamma_{X}=G^{2}\frac{v^{2}+a^{2}}{4M}d{\rm Lips}(P;q_{i},N)(\omega+\hat{\omega}+(H_{\mu}+\hat{H_{\mu}})s^{\mu})
\label{eq:diff}
\end{equation}

With new {\tt TAUOLA} version modification of  hadronic currents 
for the  semileptonic decays
should pose no problem. User can focus on physics aspects and work e.g. in  {\tt C++} environment\footnote{
Although all arrangements were prepared with {\tt C++} in mind, 
user can work in  other programming language instead. }.
Anomalous decays are not covered by the above scheme.
For those, whole matrix element may need to be prepared and ported to {\tt TAUOLA}.
This is no harder than modifications or replacements of the
hadronic currents, details will be shown later in this paper.

\section{Data structures and wrappers}\label{sec:data}

Until now, many things were hard-coded in {\tt FORTRAN} routines. Our intention was to make explicit
separation into modules: fixed multiplicity phase space generators,
hadronic currents, ME calculators, initialization and interface to event records.
At the same time we aimed to make changes in the code as small as possible.
Our present work is a follow-up on the work started in \cite{Was:2000st}
and continued in \cite{Was:2004dg}, which became a technical start for the
hadronic current parametrizations used by BaBar collaboration. Our present
arrangements  profit from the opportunities available in mixed {\tt F77/C++}
environment.

Our aim was to structurize program in the object-oriented manner but at the same time
retain the use of static data structures instead of switching to dynamic memory allocations
to preserve performance and minimize memory usage.
Such arrangement is a step towards introduction of parallelization methods
while still facilitating backward-compatibility checks.

The first step of our work was to prepare {\tt FORTRAN common blocks} designated
to keep all information regarding the generation process. The changes
introduced to the code are in {\tt FORTRAN}, but as a result,
each segment can be translated into {\tt C++} independently.

We use specialized routines to generate
phase-space for final states of fixed multiplicity.
This is the reason why the decay channels are grouped accordingly to multiplicity.
The constants {\tt NMi}, where {\tt i = 1..5}, denote number of memory slots
reserved for channels of final states with multiplicity {\tt i+1}.
For {\tt NM6}, this convention is overruled. This group collects decay channels
with final states of any multiplicity, but for these decay channels an
iterative phase space generator is used and as a consequence 
approximations on matrix element must be imposed.
Two additional  variables should be mentioned before data structures are 
explained.
The {\tt NLT=2}  denotes number of leptonic $\tau$ decay channels. The
{\tt NMODE} denotes the total number of memory slots prepared for non-SM or non-leptonic decay channels.
It's value is equal to {\tt NM1+NM2+NM3+NM4+NM5+NM6}.
In the source code provided with this paper {\tt NMODE=196}.
If needed, this value can be increased up to {\tt 500-NLT}.

\subsection{Data structures}

First, let us present these data structures
that, once initialized, usually remain constant during the remainder of the 
program execution%
\footnote{Note that it is possible to change channel branching probability dynamically during
program execution. It is also possible, however undesireable and error-prone,
to change the decay channels definition more than once during program execution.}.
\begin{verbatim}
extern "C" stuct taubra_ {
    float GAMPRT[500];
    int JLIST[500];
    int NCHAN;
};
\end{verbatim}

\begin{description}
\item{\tt GAMPRT[]} \\
The branching ratios used to define probabilities
with which choice of particular decay channel is made\footnote{
It can be set with any non-negative numbers.
User should ensure correct proportions for channels.
For example, if user generates two decay modes only,
one of which is 100-times more likely to occur,
it makes no difference for the program if respective entries in
{\tt GAMPRT} are set to 1. and 100. or 0.001 and 0.1.}.

\item{\tt JLIST[]} \\
The channel number assigned within {\tt FORTRAN} part of the code.
Usually {\tt JLIST(i) == i, where i=1,...,500}.

\item{\tt NCHAN = NMODE + NLT} \\
Number of all available decay channels.
 Parameters {\tt NLT, NMODE, NM1, NM2, NM3, NM4, NM5, NM6}
are initialized in two places {\tt TAUDCDsize.inc} and in {\tt TauolaStructs.h}.
It has to remain consistent whenever changes would be introduced.

\end{description}

\begin{verbatim}
extern "C" struct metyp_ {
    int KEY0[NLT];
    int KEY1[NM1];
    int KEY2[NM2];
    int KEY3[NM3];
    int KEY4[NM4];
    int KEY5[NM5];
    int KEY6[NM6];
};
\end{verbatim}

In this  structure information on  the type of the matrix elements
used by each decay channel is stored.
As usual, decay channels are grouped by multiplicity.
The following types of matrix element calculation are possible:
\begin{description}
\item 0 - channel is not initialized,
\item 1 - constant matrix element (flat phase space),
\item 2 - default {\tt TAUOLA} matrix element and current,
\item 3 - default  {\tt TAUOLA} matrix element,  but for case when one, stable particle of spin$>$0 
in final state%
\footnote{In practice, it is used for $\tau \to \pi^-\pi^0\gamma$ decay channel only.}
is present,
\item 4 - default matrix element but with user defined hadronic current,
\item 5 - user defined matrix element.
\end{description}

\begin{verbatim}
extern "C" struct taudcd_ {
    int  IDFFIN[NMODE][9];
    int  MULPIK[NMODE];
    char NAMES [NMODE][31];
};
\end{verbatim}

\begin{description}
\item{\tt NAMES[i][31]} \\
List of names for the decay modes to be used for the output printouts.
Up to 31 letters can be used to describe each decay channel%
\footnote{Note that variable {\tt NAMES} is defined in {\tt FORTRAN} as a variable
of type {\tt CHARACTER*31}. As such, it is not a null-terminated C string.}.

\item {\tt MULPIK[i]} \\
Multiplicity of stable decay products ($\nu_\tau$ excluded). An overall
multiplicity for the channel {\tt i} is thus {\tt MULPIK[i]+1}.

\item {\tt IDFFIN[i][j]} \\
Identifiers of the consecutive stable decay products (except $\nu_\tau$).
Entries for {\tt j $\ge$ MULPIK[i]+1} should be  set to zero.
An exception is for the anomalous decay channels where $\nu_\tau$ is absent.
In such cases non-zero {\tt IDFFIN[i][j]} for {\tt j = 3} or {\tt j = 4} denotes
identifier for the particle to be placed into decay final state instead of $\nu_\tau$.
Entry for {\tt j = 3} is to be used when identifier is to remain the same for $\tau^-$
and $\tau^+$. Entry for {\tt j = 4} should be used when the sign should differ%
\footnote{Note that this arrangement implies, that 
anomalous, neutrinoless channels can be initialized only
for 3 or 2 particles final states. We think it is sufficient
for foreseeable future.}.
\end{description}

\begin{verbatim}
extern "C" struct sampl2_ {
    float PROB1[NM2];
    float PROB2[NM2];
    float AM2[NM2];
    float GAM2[NM2];
    float AM3[NM2];
    float GAM3[NM2];
};

extern "C" struct sampl3_ {
    float PROB1[NM3];
    float PROB2[NM3];
    float AMRX[NM3];
    float GAMRX[NM3];
    float AMRA[NM3];
    float GAMRA[NM3];
    float AMRB[NM3];
    float GAMRB[NM3];
};

extern "C" struct sampl4_ {
    float PROB1[NM4];
    float PROB2[NM4];
    float AMRX[NM4];
    float GAMRX[NM4];
    float AMRA[NM4];
    float GAMRA[NM4];
};

extern "C" struct sampl5_ {
    float PROBa2[NM5];
    float PROBOM[NM5];
    float ama2[NM5];
    float gama2[NM5];
    float AMOM[NM5];
    float GAMOM[NM5];
};
\end{verbatim}

The structures collect parameters for the presamplers and  channels 
of multiplicity corresponding respectively from 2 to 5 hadrons 
in the final state.
Those parameters are  probabilities for generation sub-channels
or coefficients, such as  
 masses and widths of intermediate peaks.
Modification of a presampler affects  efficiency of generation only.
It may be necessary if substantially modified hadronic current is used,
for example,  if user plans to check the code in  intermediate resonances
 narrow width limit.

\subsection{Wrappers}\label{sec:wrappers}

Let us document functions that have to be written by the user
when an interface to the new software structure is introduced%
\footnote{An implementation of these functions have been presented
in {\tt tauola-c/channel\_wrappers.c} file.
This implementation demonstrates how parameters of 
the functions for calculation of matrix element or hadronic currents
should be used. If the {\tt C++} interface is to be removed, e.g. as the first 
step of writing another one, possibly to different laguage, use of code of {\tt tauola-no-c/ }
directory, instead of {\tt tauola-c} may be useful.}.
These wrappers are used to pass parameters from {\tt FORTRAN}
part of the code to the user-provided functions.
That is why the format of these functions must be strictly followed.

Each function  name consist of name of the current (or matrix element) as used in older
pure {\tt FORTRAN} versions of {\tt TAUOLA} but with {\tt $\_$wrap} added at the end.  
For example, wrapper for user routine to replace {\tt DAM2PI} function is declared as: \\
{\small
\begin{verbatim}
extern "C" void dam2pi_wrap_
          (int *MNUM, float *PT, float *PN, float *PIM1, float *PIM2,
           float *AMPLIT, float *HV)
\end{verbatim}
}

These wrappers depend on multiplicity of final states
but in most cases are prepared accordingly to the same scheme.
The exceptions are listed at the end of this Section.

Note that such  wrappers are executed only if user have registered a custom channel 
(with matrix element or hadronic current, i.e. {\tt KEY2[mnum]}=4 or 5)
for particular multiplicity and particular {\tt mnum} 
({\tt mnum} denotes position on the sub-list of decays corresponding to the multiplicity). 

The {\tt FORTRAN common blocs} are not sufficient to describe initialization
of {\tt TAUOLA}. The table, like  {\tt Tauolapp::taubra$\_$userChannels[channel]} of our
 interface, described in Appendix~\ref{app:interface_example} is needed. 
Let us recall essential points only, leaving description of the Interface to the Appendix.
The function {\tt RegisterChannel} of this interface can be used. 
The position on the list of all channels is calculated from the position 
of channels with two scalars in final state {\tt MNUM} as {\tt NM4+NM5+NM6+NM3+mnum}.
Then, a pointer to the user routine is taken from the table: \\
{\tt Tauolapp::taubra$\_$userChannels[channel]}. \\
This pointer is set to user routine executed in place of function: \\
{\tt xsec(pt,pn,pim1,pim2,amplit,hv)} \\
If for some reason this pointer is not set, program will exit with an error: \\
{\small \tt ERROR: dam2pi$\_$wrap: pointer to xsec for channel 3 (mnum 3) not set!"} \\

The remaining part of  each function-wrapper is devoted to debugging.
It will activate if any of the calculated quantities, matrix element squared or
polarimetric vector is of a {\tt NaN} value.

The following subsubsections list all wrappers used in the project. We use the {\tt C/C++} declaration
to describe these wrappers as it is less intuitive than the original {\tt FORTRAN} declaration,
which can be deducted much easier.
\subsubsection{Wrappers for squared Matrix Elements}

\begin{description}
\item {\tt extern "C" void dam1pi$\_$wrap$\_$
\\ (int *MNUM, float *PNU, float *AMF0, float *PKK, float *AMF1, float *GAMM, float *HV) } \\
\item {\tt extern "C" void dam2pi$\_$wrap$\_$
\\ (int *MNUM, float *PT, float *PN, float *PIM1, float *PIM2, float *AMPLIT, float *HV) } \\
\item {\tt extern "C" void dam3pi$\_$wrap$\_$
\\ (int *MNUM, float *PT, float *PN, float *PIM1, float *PIM2, float *PIM3, float *AMPLIT, float *HV) }  \\
\item {\tt extern "C" void dam4pi$\_$wrap$\_$
\\ (int *MNUM, float *PT, float *PN, float *PIM1, float *PIM2, float *PIM3, float *PIM4, float *AMPLIT, float *HV) } \\
\item {\tt extern "C" void dam5pi$\_$wrap$\_$
\\ (int *MNUM, float *PT, float *PN, float *PIM1, float *PIM2, float *PIM3, float *PIM4, float *PIM5, float *AMPLIT, float *HV) }
\end{description}

\subsubsection{Wrappers for Hadronic Currents}

\begin{description}
\item {\tt extern "C" void curr2$\_$wrap$\_$
\\ (int *MNUM, float *PIM1, float *PIM2, complex *HADCUR) } \\
\item {\tt extern "C" void curr3pi$\_$wrap$\_$
\\ (int *MNUM, float *PIM1, float *PIM2, float *PIM3, complex *HADCUR) } \\
\item {\tt extern "C" void curr4$\_$wrap$\_$
\\ (int *MNUM, float *PIM1, float *PIM2, float *PIM3, float *PIM4, complex *HADCUR) } \\
\item {\tt extern "C" void curr5$\_$wrap$\_$
\\ (int *MNUM, float *PIM1, float *PIM2, float *PIM3, float *PIM4, float *PIM5, complex *HADCUR) } \\
\end{description}

\subsubsection{Special cases}

\begin{description}
\item {\tt extern "C" void dampry$\_$wrap$\_$
\\ (int *ITDKRC, double *XK0DEC, double *XK, double *XA, double *QP, double *XN, double *AMPLIT, double *HV) } \\
\hskip 1 cm Matrix element calculation for leptonic decays, including
possibly 4-momentum for bremsstrahlung photon
and its phase space limit {\tt XK0DEC}.

\item {\tt extern "C" float sigee$\_$wrap$\_$(float *AMX2, int *MNUM) } \\
In some decays approximation is used. Only distribution of invariant mass for all
hadrons is generated accordingly to function {\tt sigee} as if it was vector state.
The final states of hadronic system fragmentation are generated accordingly to flat phase space.

\item {\tt extern "C" float fconst$\_$wrap$\_$(int *MNUM) } \\
In this case no matrix element is used at all. Only overall constant is passed
to generation. Flat phase space (all mass effects included) distribution is generated.
\end{description}

\subsection{Generation monitoring variables}

All structures presented so far, were for  variables initialized
at the vey beginning of initialization and left unchanged over the whole generation run.
Wrappers are not expected to store information on the whole sample as well.
This Subsection presents the variables  used to access information which is 
continuously updated during the program run.
Information on generation weights is stored in local variables of 
routine {\tt DADNEW} for each semileptonic decay
separately, and in routines {\tt DADMEL} and {\tt DADMMU} for leptonic decays:
\begin{verbatim}
      REAL*4 WTMAX(NMODE)
      REAL*8              SWT(NMODE),SSWT(NMODE)
      INTEGER*8 NEVRAW(NMODE),NEVOVR(NMODE),NEVACC(NMODE)
\end{verbatim}
The variables denote:
\begin{description}
\item{\tt WTMAX} - maximum weight for the given channel,
\item{\tt SWT} - sum of all event weights for the given channel,
\item{\tt SSWT} - sum of all event weights squared for the given channel,
\item{\tt NEVRAW} - number of raw generated events for the given channel,
\item{\tt NEVOVR} - number of evens of weights larger than the maximum estimated at the initialization for the given channel,
\item{\tt NEVACC} - number of accepted events for the given channel.
\end{description}

Further variables regarding the program run are stored in {\tt common block TAUBMC}:

\begin{verbatim}
       COMMON / TAUBMC / GAMPMC(500),GAMPER(500),NEVDEC(500)
\end{verbatim}

The variables denote:

\begin{description}
\item{\tt GAMPMC(500)} - width for the given decayc channel calculated from the Monte Carlo generation,
\item{\tt GAMPER(500)} - error of the {\tt GAMPMC},
\item{\tt NEVDEC(500)} - number of generated decays.
\end{description}

Parts of the code and variables of the present Subsection will have to be adjusted
at later steps of the program evolution, but as these parts do not affect the
use of the presented methods for matrix element replacement we left them
in {\tt FORTRAN}. 

This setup has  to be modified
if a multi-threaded parallelization process is to be used.
At this moment it does not seem to be necessary, because with the present day
matrix elements {\tt TAUOLA} is sufficiently fast.
Furthermore, applying parallelization solution as used by {\tt KK} Monte Carlo is
straightforward. It would only require to store run information, described 
in the present Subsection, into histograms.

\section{Default {\tt TAUOLA} initialization}\label{sec:initializations}

The main purpose of the present paper is description of the means
for user alterations of hadronic currents to be used in $\tau$ decays. However,
these modifications have to base on the default initialization of {\tt TAUOLA}.
In this Section we describe  such
default initialization available for the user prior to his main work.

For present distribution of {\tt TAUOLA} we decided that the initialization 
compatible with the one used by BaBar collaboration, from now on will be used as default.
It contains currents which even with very large statistics reproduce
results of BaBar collaboration version of {\tt TAUOLA}. This is the version
used by collaboration for their default simulations\footnote{
We have no detailed knowledge when and which other currents
parametrizations were used for specific measurements/studies. Nonetheless
we think that this version deserves archivization.}.
We have compared results of our new initialization of {\tt TAUOLA} with BaBar collaboration production files. 
We have reproduced BaBar setup with a great detail to the best of our knowledge. 
However, we still have missed fine-tuning of  minimum photon energy 
for {\tt PHOTOS}~\cite{Barberio:1993qi,Golonka:2005pn} 
and {\tt TAUOLA} for generation of bremsstrahlung in decays.
Therefore we did not have reproduced exactly results of BaBar production files.
However, the differences were minuscule. 
Agreement was checked with {\tt MC-TESTER} \cite{Davidson:2008ma}
with samples of 1 617 945 000 $\tau$ decays.
The {\tt MC-TESTER}-based tests accounted for over 133 decay channels
(including those with multiple photons in final state generated by {\tt PHOTOS}).
Dissimilarity coefficients calculated by {\tt MC-TESTER} (see Ref.~\cite{Golonka200439} Section 6.1)
were: T1=0.033235 \%, T2=0.058574 \%.
Further details on the test and how to reproduce its results are given in Appendix~\ref{subsec:valid}.

\subsection{Extensions}

In principle, there are several extensions that can be applied to this initialization.
The distribution tarball includes e.g.
{\tt RChL} currents for 3 pions \cite{Nugent:2013hxa}, 
Novosibirsk currents for 4 pions~\cite{bondar:2002mw}, 
and 5 pion currents described in~\cite{kuhn:2006nw}.   
Those additional currents are of secoundary importance so we will not describe them here.
Information on usage is given in respective publications, {\tt README} files and comments in the code.

We would like to point out that this distribution does not provide
CLEO initialization of {\tt TAUOLA} used since Ref.~\cite{Was:2000st},
even though the code for this parametrization is not removed but left inactive. 
That is the reason why some comments in the code are outdated.
We feel that since the present version is prepared to start completely
new initialization, archiving BaBar initialization and setting it as the default
brings more benefit than retaining CLEO and/or other initializations used so far. 
If the need arises user may adapt the existing initialization to match CLEO setup
at his own discretion.
To decide whether to remove corresponding code or make it available for the user
will rely on discussions with community of users.

\section{Example of the user-defined hadronic current} \label{sec:Exampleinst}

In this Section we present an example of a user-defined matrix element, which we 
exploit to extend the default setup of {\tt TAUOLA} described in previous Section.
For this purpose, anomalous channels are good candidates because of large
interest from the users. Moreover, these channels are not as challenging for fitting
as the other ones. Their analytic form is simple. Present Section describes
the physics aspects of these matrix elements. The distribution tarball contains
a {\tt demo-lfv} directory demonstrating how such channels can be added into
{\tt TAUOLA} using the new setup.
A practical example is given in Appendix~\ref{app:install}.

\subsection{Effective field theory approach as applied to $\tmmm$ decay}
\label{subchap:EQFT}

CLFV processes resulting from BSM theories can be described in a model-independent
way in terms of operators. If new physics exist at a mass scale $\Lambda$,
it can manifest itself at an electroweak scale in the form of a higher order operators which,
however, do not spoil the $SU(2)_L \times U(1)_Y$ symmetry. 
In the EFT approach the right-handed singlets are written as the following isospin doublets~\cite{turczyk}:

\begin{equation}
\label{eq:EW_semidoublet}
R_e=\dfrac{1-\gamma_5}{2}{0 \choose \psi_{e}},~ R_{\Pmu}=\dfrac{1-\gamma_5}{2}{0 \choose \psi_{\mu}},~
R_{\Ptau}=\dfrac{1-\gamma_5}{2}{0 \choose \psi_{\tau}}.
\end{equation}
Taking into account Eq.~ \ref{eq:EW_semidoublet} and the matrix of Higgs fields 
from~\cite{turczyk}, one can derive the following relevant dimension six operators:

\begin{align}
O_1 &=(\bar{L} \gamma_{\Pmu} L)(\bar{L}\gamma^{\Pmu}L), \label{eq:operators:01}\\
O_2 &=(\bar{L} \Ptau^a \gamma_{\Pmu} L)(\bar{L} \Ptau^a\gamma^{\Pmu}L), \label{eq:operators:02} \\
O_3 &=(\bar{R}  \gamma_\mu R)(\bar{R} \gamma^{\mu}R), \label{eq:operators:03} \\
O_4 &=(\bar{R}  \gamma_\mu R)(\bar{L} \gamma^{\mu}L),  \label{eq:operators:04}\\
R_1 &=g'(\bar{L} H \sigma_{\mu \nu}R)B^{\mu \nu},  \label{eq:operators:R1} \\
R_2 &=g(\bar{L} \tau^{a} H \sigma_{\mu \nu}R)W^{\mu \nu}. \label{eq:operators:R2}
\end{align}

As defined above, $B_{\mu \nu}$ and $W_{\mu \nu, a}$ are the electroweak gauge fields,
$g$ and $g'$ are the coupling constants of $SU(2)_L$ and $U(1)_Y$, $H$ denotes
the matrix of Higgs fields, $L(R)$ are the left(right)-handed fields
and $\sigma^{\mu,\nu}=\dfrac{i}{4}\left[\gamma^{\mu}, \gamma^{\nu} \right]$.
According to S.~Turczyk et. al.~\cite{turczyk2}, higher order operators
are suppressed by small lepton Yukawa couplings. We will not consider them in this paper.
In the effective field theory the most general Hamiltonian that describes the discussed
process is formed as the sum of the operators  \ref{eq:operators:01} to \ref{eq:operators:R2}.
For the process $\tmmm$ the operators $O_1$ and $O_2$ are identical
after projecting them on charged leptons. The $O_3$ corresponds to a purely
right-handed current and is completely analogous to $O_1$. The $O_4$ operator corresponds to the mix four fermion operator. For radiative operators $R_1$ and $R_2$, the latter is suppressed by small Yukawa coupling
of $\Ptau$. Only  photonic operator $R_1$ is relevant.

\begin{figure}[htb]
\begin{center}
\subfloat[Simulated Dalitz distr. for Eq.~\ref{eq:LL}.]{\includegraphics[angle=-0,width=0.42\textwidth]{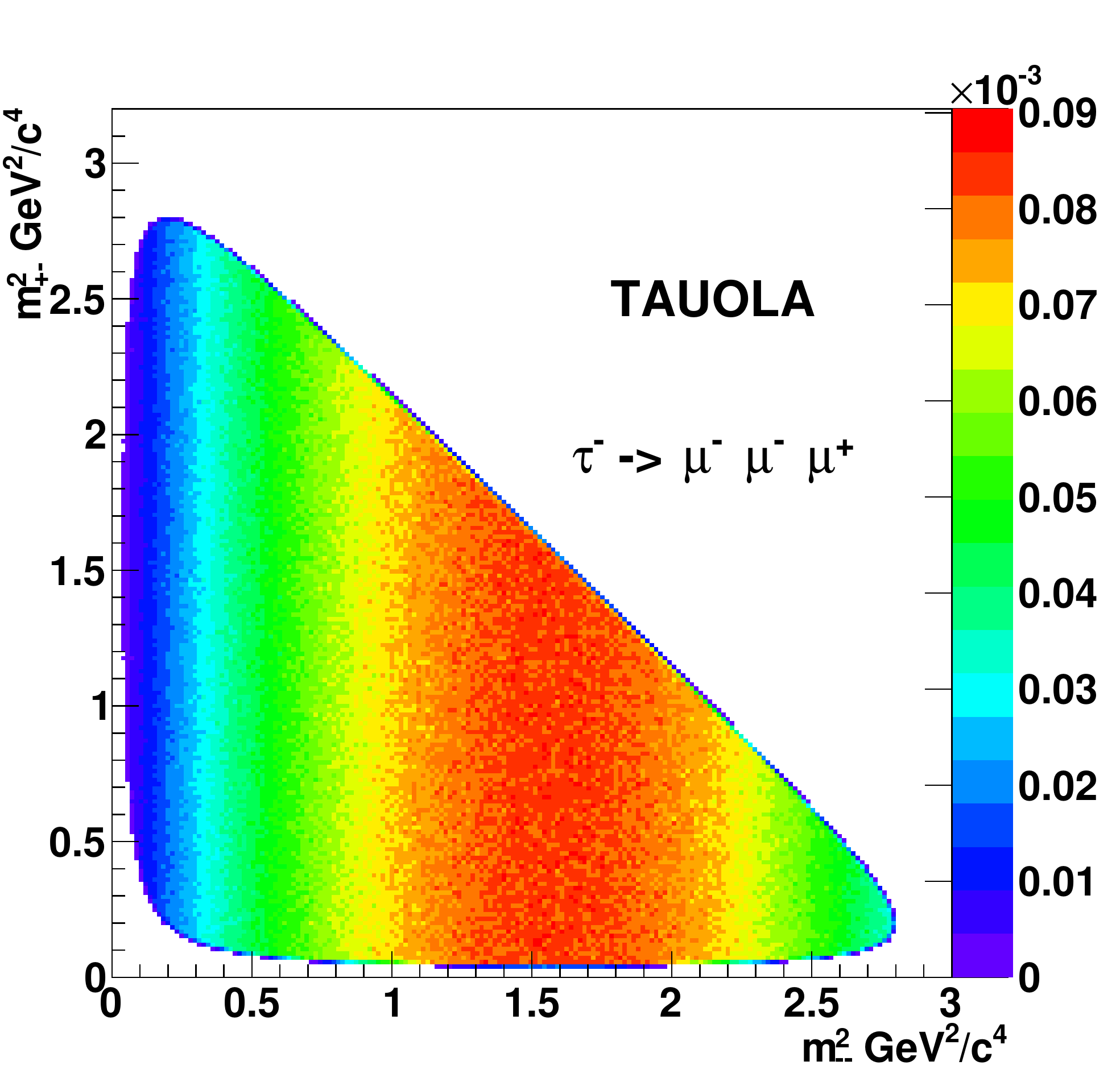}}
\subfloat[Simulated Dalitz distr. for Eq.~\ref{eq:LR}.]{\includegraphics[angle=-0,width=0.42\textwidth]{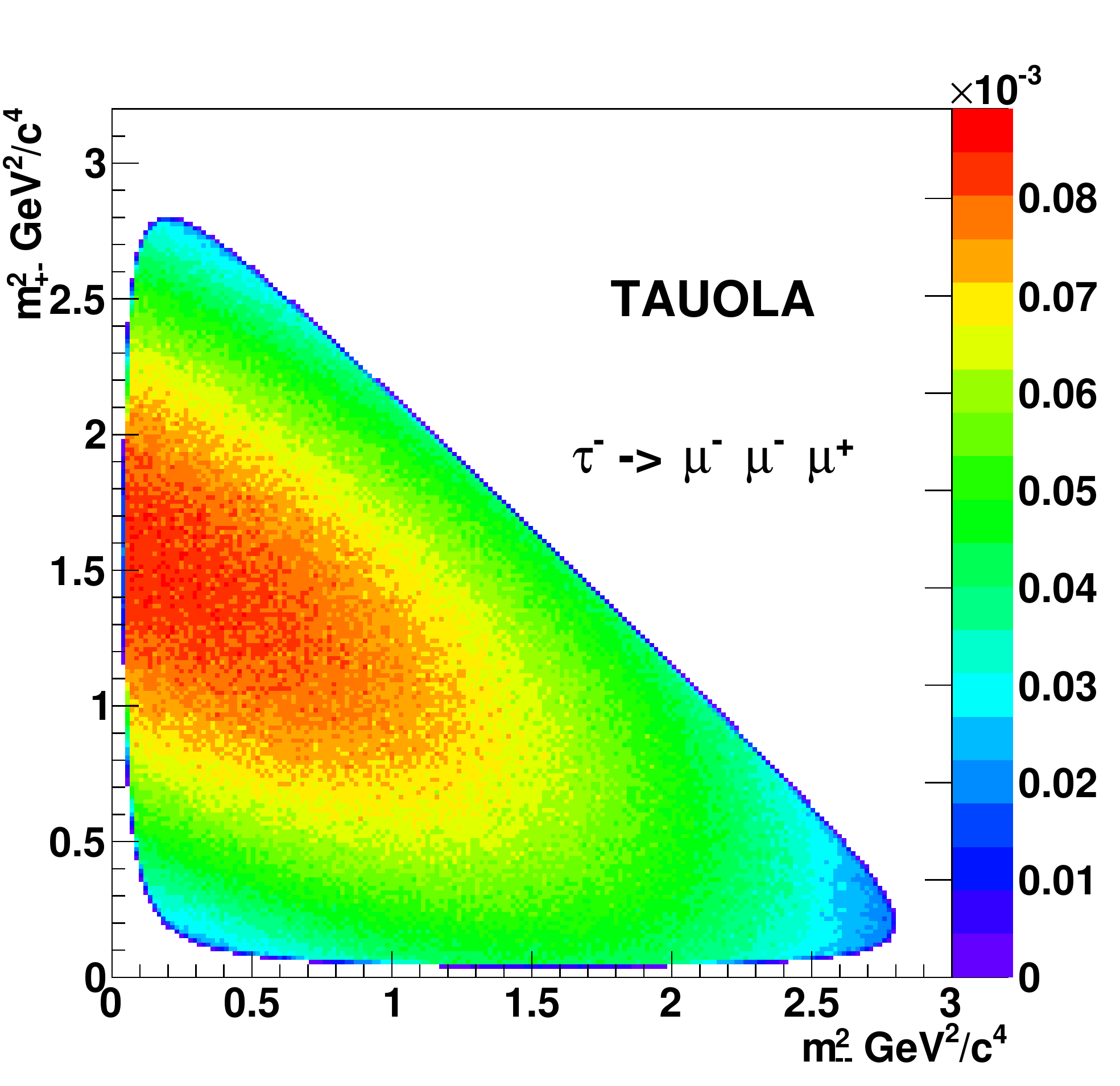}}\\               
\subfloat[Simulated Dalitz distr. for Eq.~\ref{eq:rad}.]{\includegraphics[angle=-00,width=0.42\textwidth]{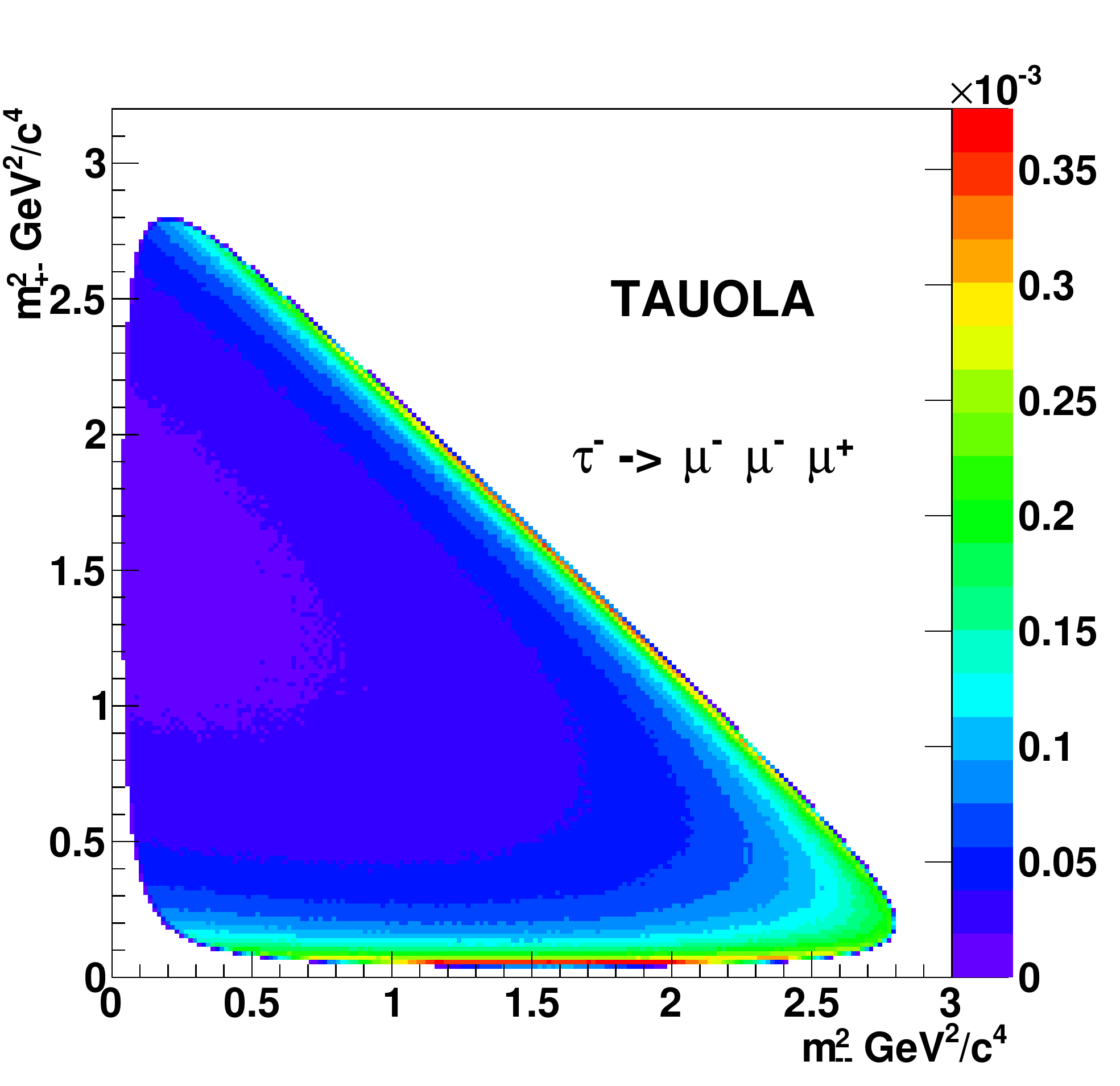}}
\subfloat[Simulated Dalitz distr. for Eq.~\ref{eq:radLL}.]{\includegraphics[angle=-0,width=0.42\textwidth]{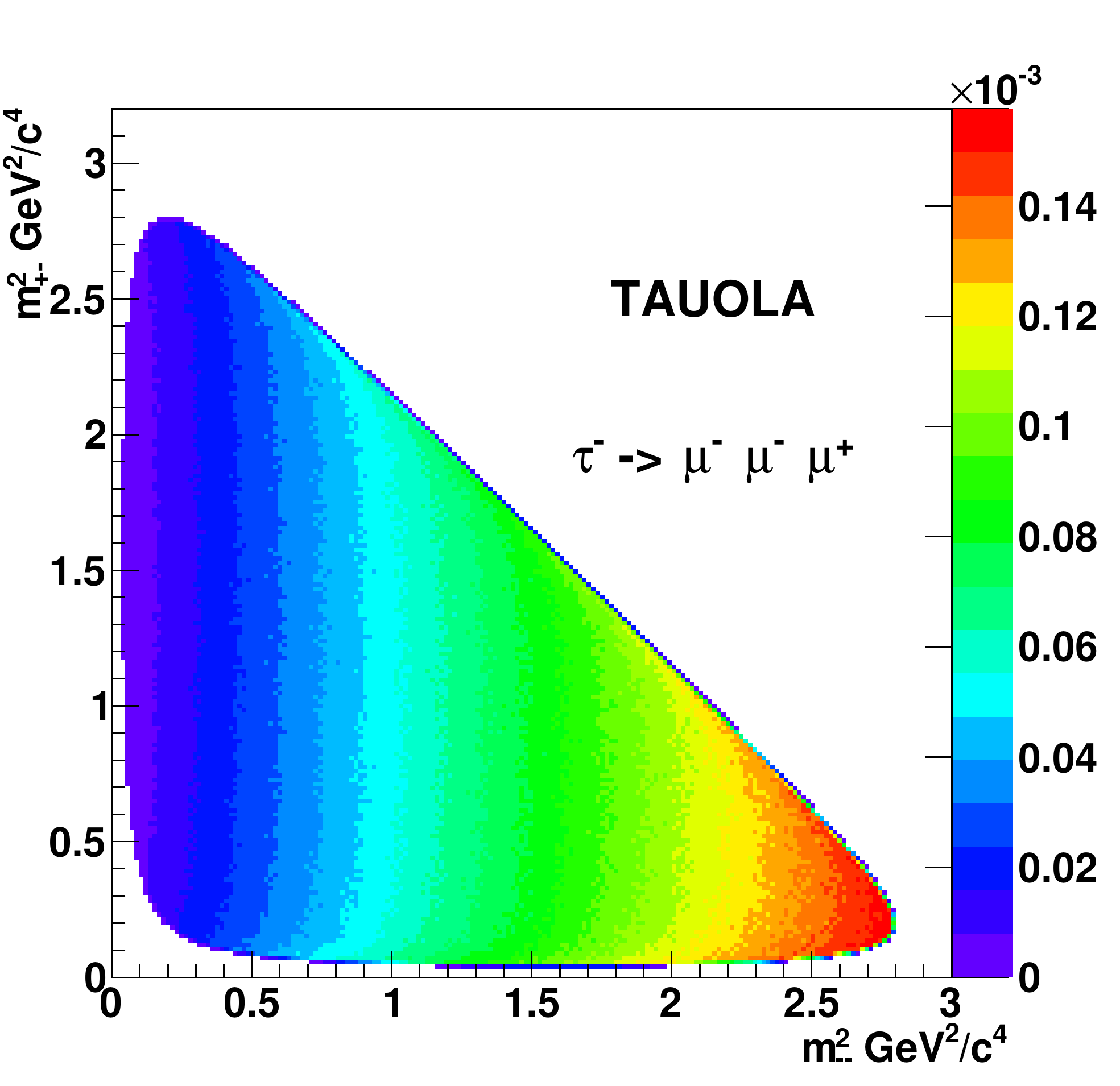}}\\
\subfloat[Simulated Dalitz distr. for Eq.~\ref{eq:radLR}.]{\includegraphics[angle=-0,width=0.42\textwidth]{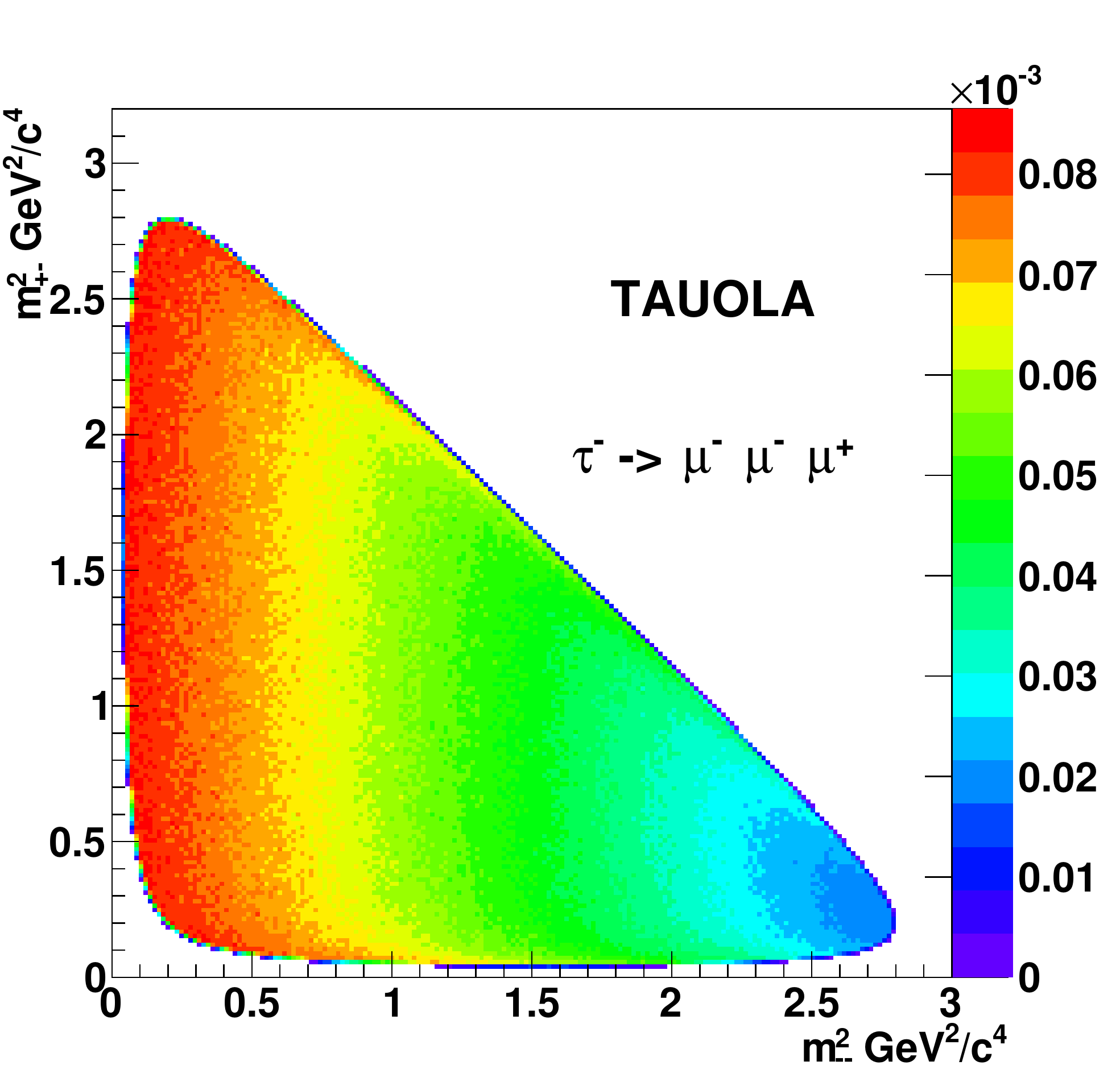}}

    \end{center}
\caption{\small Dalitz distributions simulated in the effective field approach for
         the five different BSM operators corresponding to different lepton chirality structures~\cite{turczyk}.
         The normalized to unit area distributions, implemented in the {\tt TAUOLA} package.}
 \label{fig:dalitz_tauola}
\end{figure}

The operators of the present paper were also interpreted in terms of the BSM operators.
The respective decay widths can be presented in the form of Dalitz distributions~\cite{PhysRev.94.1046},
which were derived in the following five cases, corresponding to different
lepton chirality structures:

\begin{itemize}
\item Four left-handed leptons ($O_1$ operator):
\end{itemize}
\begin{equation}
\label{eq:LL}
\dfrac{d^2\Gamma_V^{(LL)(LL)}}{dm_{23}^2dm_{12}^2} =\dfrac{\left| g_V^{(L_{\Pmu}L^{\Ptau})(L_{\Pmu}L^{\Pmu})} \right|^2}{\Lambda^4} \dfrac{(m^2_{\Ptau}- m_{\Pmu}^2)^2 - (2m_{12}^2 -m_{\Ptau}^2 - 3m_{\Pmu}^2)^2 }{256 \pi^3 m_{\Ptau}^3}.
\end{equation}
\begin{itemize}
\item Two left-handed, two right-handed leptons ($O_4$ operator):
\end{itemize}
\begin{align}
\label{eq:LR}
\dfrac{d^2\Gamma_V^{(LL)(RR)}}{dm_{23}^2dm_{12}^2} & =  \dfrac{\left| g_V^{(L_{\Pmu}L^{\Ptau})(R_{\Pmu}R^{\Pmu})} \right|^2}{\Lambda^4} \left[ \dfrac{(m^2_{\Ptau}- m_{\Pmu}^2)^2 - 4 m_{\Pmu}^2 (m_{\Ptau}^2 +m_{\Pmu}^2 - m_{12}^2) }{512 \pi^3 m_{\Ptau}^3 } \right.  \nonumber \\
 &\qquad {} \left. - \dfrac{(2m_{13}^2 - m_{\Ptau}^2 -3m_{\Pmu}^2)^2   + (2m_{23}^2-m_{\Ptau}^2 -3m_{\Pmu}^2)^2 }{1024 \pi^3 m_{\Ptau}^3}  \right].
\end{align}
\begin{itemize}
\item Radiative right-handed $\tau$ leptons ($R_1$ operator):
\end{itemize}
\begin{align}
\label{eq:rad}
 \dfrac{d^2\Gamma_{rad}^{(LR)}}{dm_{23}^2dm_{12}^2} & = \alpha^2_{em} \dfrac{\left| g_{rad}^{(L_{\Pmu}R^{\Ptau})} \right|^2 \nu^{2}}{\Lambda^4} \left[  \dfrac{ m_{\Pmu}^2 (m_{\Ptau}^2 -m_{\Pmu}^2)^2} {128 \pi^3 m_{\Ptau}^3}(\dfrac{1}{m_{13}^4}+\dfrac{1}{m_{23}^4}) \right.   \nonumber \\
&\qquad {} \left. + \dfrac{m_{\Pmu}^2(m_{\Ptau}^4-3m_{\Ptau}^2 m_{\Pmu}^2 +2m_{\Pmu}^2)} { 128 \pi^3 m_{\Ptau}^3 m_{23}^2 m_{13}^{2}}  + \dfrac{2m_{12}^2-3m_{\Pmu}^2}{ 128 \pi^3 m_{\Ptau}^3} \right.   \nonumber \\
&\qquad {} \left. +  \dfrac{(m_{13}^2 +m_{23}^2)(m_{12}^4+m_{13}^4+m_{23}^4-6m_{\Pmu}^2(m_{\Pmu}^2+m_{\Ptau}^2)) }{256 \pi^3 m_{\Ptau}^3   m_{23}^2  m_{13}^{2}} \right].
\end{align}
\begin{itemize}
\item Interference between $O_1$ and $R_1$:
\end{itemize}
\begin{align}
\label{eq:radLL}
\dfrac{d^2\Gamma_{mix}^{(LL)(RR)}}{dm_{23}^2dm_{12}^2} & = \alpha^2_{em} \dfrac{2 \nu Re \left[ g_{V}^{(L_{\Pmu}L^{\Ptau})(L_{\Pmu}L^{\Pmu})} g_{rad}^{*L_{\Pmu}R^{\Ptau}} \right] }{\Lambda^4} \left[ \dfrac{m_{12}^2-3m_{\Pmu}^2}{64 \pi^3 m_{\Ptau}^2} + \right.   \nonumber \\
&\qquad {} \left.  \dfrac{m_{\Pmu}^2(m_{\tau}^2-m_{\Pmu})^2(m_{13}^2+m_{23}^2)}{128 \pi^3 m_{\Ptau}^2 m_{23}^2 m_{13}^{2}} \right].
\end{align}
\begin{itemize}
\item Interference between $O_4$ and $R_1$:
\end{itemize}
\begin{align}
\label{eq:radLR}
\dfrac{d^2\Gamma_{rad}^{(LL)(RR)}}{dm_{23}^2dm_{12}^2}&  = \alpha_{em} \dfrac{2 \nu Re \left[ g_{V}^{(L_{\Pmu}L^{\Ptau})(R_{\Pmu}R^{\Pmu})} g_{rad}^{*L_{\Pmu}R^{\Ptau}} \right] }{\Lambda^4} \left[ \dfrac{m_{\Ptau}^2 -m_{12}^2 -3m_{\Pmu}^2}{256 \pi^3 m_{\Ptau}^2} +  \right.   \nonumber \\
&\qquad {} \left.   \dfrac{m_{\Pmu}^2(m_{\Ptau}^2-m_{\Pmu}^2)(m_{13}^2+m_{23}^2)}{256 \pi^3 m_{\Ptau}^2  m_{23}^2 m_{13}^{2}} \right].
\end{align}
In distributions \ref{eq:LL} - \ref{eq:radLR} the following dimuon masses are defined:

\begin{align}
m_{--}^2= m_{12}^2=( p_{\Pmu^-} +p'_{\Pmu^-} )^2, \qquad
m_{+-}^2= m_{23}^2=( p'_{\Pmu-} +p_{\Pmu^+} )^2,  \nonumber \\
m_{13}^2=m_{\Ptau}^2 + 3m_{\Pmu}^2 - m_{--}^2 - m_{+-}^2, \qquad  
\end{align}

and $m_{\ell}$ is the mass of corresponding lepton, $g_{V}$ is the corresponding coupling constant
 and $\nu$ is the element from the Higgs matrix. The Dalitz dostributions can be found 
in~Fig.~\ref{fig:dalitz_tauola}, we were able to reproduce them with the help of {\tt TAUOLA} 
Monte Carlo.

Note that in contrary to phase-space parametrization of Eq.~\ref{eq:ps},
in case of Dalitz parametrization, $dm^2_{23}dm^2_{12}$, no phase space,
kinematic-dependent, contribution to Jacobians 
appear. That is why, right hand sides of Eqs.~\ref{eq:LL} to \ref{eq:radLR} are
proportional, up to mumerical overall constant to matrix element squared
and averaged/summed over the spin.  
This is true, because  no dependence on 
angular variables is present in these formulae. We have implemented these formulae into our program, neglecting sometimes terms of order of 
$\sim \frac{m_\mu^2}{m_\tau^2} < 0.01$.

\clearpage

\section{Summary}
\label{sec:summary}
The purpose of the present paper was to present an installation package
for essentially {\tt FORTRAN} update of  $\tau$ leptons
decay library  {\tt TAUOLA}.
The package is ready to be merged with
{\tt Tauola Universal Interface} \cite{Davidson:2010rw} and KKMC \cite{Jadach:1999vf}: 
necessary patches are provided.
The distribution corresponds nearly in all details
to the version presented at TAU14 conference
\cite{Was:2014zma}; minor modifications are explicitly listed.

In the code distributed with the present  paper,
 maximum number of $\tau$ decay channels is  500, 
most of them are left not-initialized and represent reserved static memory only.
Changes introduced into {\tt FORTRAN} part of {\tt TAUOLA} have been described in a great detail.
All initialization information has been moved to {\tt FORTRAN common blocks} so that they can be accessed from other parts of the code, for example as C {\tt struct}s.

The default initialization of the package provides hadronic currents
and corresponding branching fractions numerically equivalent
to the ones used by the BaBar collaboration for their default simulations.
Extended set of decay channels with anomalous nature was also prepared.
For the anomalous decay channels of 2 and 3 decay products $\nu_\tau$-less channels can be introduced.

The backward compatibility to CLEO initialization was not assured,
but it is rather easy  to recover if the need would arise. 
The corresponding {\tt FORTRAN } code is not removed yet.
This distribution completes the first step towards the migration of the whole project to {\tt C++}
and towards perparation of convenient structure for the parallelization.
Further steps will require collaboration with the program users, in particular
with the authors of fitting programs of the  future experiments.
Also, at this moment, we can not decide if the {\tt FORTRAN} code for
the variants of hadronic current parametrization can be removed.

\centerline {\bf Acknowledgments}

This project  was partially supported by the funds of Polish National Science
Center under decision  UMO-2014/15/B\-/ST2/00049.

M.~Chrzaszcz is grateful for support of the Polish National Science Center
under the ''Sonata'' grant: UMO-2015/17/D/ST2/03532.

\appendix

\section{Example of the interface implementation}\label{app:interface_example}

The main goal of new {\tt TAUOLA} version was to prepare
{\tt FORTRAN} common blocks to facilitate addition of new currents and
modification of existing ones%
\footnote{In principle, there is no need for any interface to modify hadronic currents
used in {\tt TAUOLA}. Nonetheless it may be convenient for some users.}
in cases when user main program is in  {\tt FORTRAN} or in {\tt C++}. 
New setup allows for
initialization or re-initialization of the currents definition at runtime.
It also makes it easier to understand in principle modular structure of 
the algorithms
used in the project.

Before describing the implementation of the interface provided with the source code
we want to point out that most of the arrangements
presented in this Section can be modified or replaced.
All of  presented methods, have to rewrite information
stored in the data structures described in Section~\ref{sec:data}.
We have prepared solutions on the basis of discussion with users,
but alternative solutions may be found to be more convenient.
The code presented in this Section may thus be replaced rather easily.

The {\tt C++} interface for adding and modifying decay channels in {\tt TAUOLA}
is defined in {\tt tauola-c/ChannelForTauolaInterface.c}. Most of the functions provided by this interface
work on the instances of the class {\tt ChannelForTauola} (defined in {\tt tauola-c/ChannelForTauola.h}).
Functions invoked from the {\tt FORTRAN} code are defined in {\tt tauola-c/channels$\_$wrappers.c}.
These functions are used by {\tt FORTRAN} code of {\tt TAUOLA} to access {\tt ChannelForTauola} objects
and in this way pass parameters further to user-initialized channels. In particular, to user routines for
matrix elements or hadronic current calculation, accessed at the execution time through pointers.
This approach allows to program in any language as long as proper wrappers are provided.
Scheme of calling functions written in other languages from {\tt FORTRAN} depends on a
particular programming language and a set of compilers used to combine these languages.

Initialization of user defined code is performed in two steps.
First, one has to call routine {\tt Tauolapp::setUserRedefinitions(RedefExample);}%
\footnote{In example {\tt demo-redefine} it has been shown how this routine can be
called from {\tt FORTRAN} via {\tt subroutine  TAUOLAREDEF} defined in {\tt demo-redefine/iniofc.c}}.
Then {\tt iniofc} routine (defined in {\tt tauola-c/channels$\_$wrappers.c})
has to be called to reinitialize {\tt TAUOLA} with new user-updated currents.

Note that useful information can be found also in comments of the header files
located in {\tt tauola-c} subdirectory and in the example programs provided
with the distribution. Most notably, an extensive example of use of the
interface described in this Appendix is available in {\tt demo-redefine}
subdirectory. File {\tt demo-redefine/iniofc.c} contains verbose comments
about available functions, their effects and possible uses.
All {\tt FORTRAN common blocks} used by the interface are declared as
external {\tt C} structures in file {\tt tauola-c/TauolaStructs.h}.
Note that in this file parameters {\tt NLT, NMODE, NM1, NM2, NM3, NM4, NM5}
and {\tt NM6} are defined. In order for the interface to work correctly,
these definitions must coincide with the definitions located in {\tt TAUDCDsize.inc}.

\subsection{Functions defined in {\tt ChannelForTauolaInterface.h}}

Let us describe functions presented in {\tt tauola-c/ChannelForTauolaInterface.h}
which form an example of such interface.
Note that, as pointed before, this solution should serve as a prototype and can be
easily replaced if the user wishes to introduce a new one. Especially if some
of the functionality of this prototype is not needed.
For example, significant part of the code in this implementation
is devoted to preventing the coding errors when redefining or adding new channels
through strict differentiation between channels of different multiplicity.
Strict type checks are applied on such channels and function pointers passed to
these channels. Because of that, our solution may be hard to read and it might
hinder understanding the main purpose of each function.

On the other hand, understanding the interface details may not be necessary to
perform basic operations and the following explanations should be enough
to make a good use of this interface.


\begin{description}
\item {\tt extern ChannelForTauola* taubra\_userChannels[500]; } \\
Holds pointers to user defined {\tt ChannelForTauola} objects.
It is a supplement to struct {\tt taubra\_} defined in {\tt tauola-c/TauolaStructs.h}.

\item {\tt extern void*             leptonChannelsMEpointers[2]; } \\
Pointers to matrix elements  for leptonic channels.

\item {\tt extern void            (*channelRedefinitionFunction)(); } \\
Pointer to tauola channel redefinition function.

\item {\tt void PrintChannelInfo(int channel);} \\
Prints current definition of a channel: channel number,
branching ratio, multiplicity of hadrons in final state,
sub-channel number, matrix element type, and identifiers of final state particles.

\item {\tt void SetUserRedefinitions(void (*function)());} \\
Sets pointer to function reinitializing Tauola channels.
This function is the only place where channels can be reinitialized.

\item {\tt int RegisterChannel(int channel, ChannelForTauola *pointer);} \\
Changes information about selected {\tt channel} or adds a new channel.
Parameter {\tt channel} defines place on the list of channels where new one is to be added.
Setting it to {\tt -1} means that new channel has to be added on the first empty slot.
Parameter {\tt pointer} is a pointer to an object defining new or modified channel.
Function returns added or modified channel number or 0 in case of failure.
If function succeeds, it takes ownership of the object pointed by the {\tt pointer}.
Same channel number cannot be registered twice. Use {\tt OverwriteChannel} instead.

\item {\tt int OverwriteChannel(int channel, ChannelForTauola *pointer);} \\
Overwrites information about selected {\tt channel} with new information
provided by the {\tt pointer}.
Parameter {\tt channel} defines place on the list of channels where new one is to be added.
Parameter {\tt pointer} is a pointer to an object defining new or modified channel.
If function succeeds, it takes ownership of the object pointed by the {\tt pointer}.
Function returns added or modified channel number or 0 in case of failure.

\item {\tt int ModifyLeptonic(int channel, int me, float br, void *xsec = NULL);}\\
Changes information about selected leptonic channel.
This functionality is by far more delicate than that of function {\tt RegisterChannel}, and
user is requested to perform detailed tests of its performance.
Parameter {\tt channel} can be chosen only as 1 or 2.
Parameter {\tt me} is a matrix element type. See {\tt README} for the meaning.
Parameter {\tt br} is a branching ratio of the selected channel.
Parameter {\tt current} is a pointer to function calculating the cross-section or current.
Pointer {\tt xsec} has to be of type {\tt DAMPRY\_POINTER\_TYPE} defined in {\tt tauola-c/ChannelForTauola.h}.
Function returns modified channel number or 0 in case of failure.

\item {\tt ChannelForTauola* GetChannel(int channel);}\\
Get channel information from {\tt FORTRAN} common block.
Returns a pointer to {\tt ChannelForTauola} object with channel information filled,
ready to register without any changes or to update and register.
Such {\tt ChannelForTauola} object can be only registered back on their original place,
unless matrix element calculation is reduced to constant (flat phase-space),
then it can be registerd on empty slot.

\item {\tt int SetPresampler2(ChannelForTauola *pointer, float prob1, float prob2, float am2, float gam2, float am3, float gam3);}\\
Set parameters used to optimize efficiency of 2-scalar mode phase space generator.
First argument is a pointer for {\tt ChannelForTauola} for which presampler parameters will be changed.
All of the following arguments are new numerical values of presampler parameters.
Note: {\tt 0<=prob1; 0<=prob2; prob1+prob2<=1}.

\item {\tt int SetPresampler3(ChannelForTauola *pointer, float prob1, float prob2, float amrx,
                   float gamrx, float amra, float gamra, float amrb, float gamrb);} \\
Set parameters used to optimize efficiency of 3-scalar mode phase space generator.
First argument is a pointer for {\tt ChannelForTauola} for which presampler parameters will be changed.
All of the following arguments are new numerical values of presampler parameters.
Note: {\tt 0<=prob1; 0<=prob2; prob1+prob2<=1}.

\item {\tt int SetPresampler4(ChannelForTauola *pointer, float prob1, float prob2, float amrx,
       float gamrx, float amra, float gamra);} \\
Set parameters used to optimize efficiency of 4-scalar mode phase space generator.
First argument is a pointer for {\tt ChannelForTauola} for which presampler parameters will be changed.
All of the following arguments are new numerical values of presampler parameters.
Note: {\tt 0<=prob1; 0<=prob2; prob1+prob2<=1;} only {\tt prob1+prob2} is used.

\item {\tt int SetPresampler5(ChannelForTauola *pointer, float proba2, float probom,
                   float ama2, float gama2, float amom, float gamom);} \\
Set parameters used to optimize efficiency of 5-scalar mode phase space generator.
First argument is a pointer for {\tt ChannelForTauola} for which presampler parameters will be changed.
All of the following arguments are new numerical values of presampler parameters.
Note {\tt 0<=proba2<=1; 0<=probom<=1}.

\end{description}

\subsection{Key functionality of the {\tt ChannelForTauola} class}

\begin{description}

\item Constructors \\
Class {\tt ChannelForTauola} provides eleven specialized constructors, one for
each type of the user functions that can be passed to {\tt TAUOLA}%
\footnote{Matrix elements for channels of multiplicity from 1 to 5, hadronic currents for channels
of multiplicity from 2 to 5, function for a leptonic current and a constant function.}. Each of these
constructors accept the following parameters: \\
\\
{\tt float br} - branching ratio \\
{\tt const vector<int> \&ki} - list of the ID of the decay products \\
{\tt string name} - name of the decay channel \\
{\tt *\_POINTER\_TYPE pointer} - pointer to user function \\
\\
The possible {\tt POINTER\_TYPE}s are listed at the top of the {\tt tauola-c/ChannelForTauola.h} file.
Initialization of parameters {\tt mulpik} and {\tt me\_type} is done accordingly
to this pointer type. For channels that are constructed without a pointer to function,
a separate constructor has been provided that allows setting {\tt mulpik} and {\tt me\_type}
manually.

\item Accessors \\
Get and set methods have been provided for all of the parameters of this class.
The most notable are the accessors for {\tt me\_type}. The {\tt me\_type} can
be set to following values:\\
\\
0 - channel is not initialized, \\
1 - channel is reduced to constant matrix element, i.e. flat phase space is used, \\
2 - default matrix element is used, \\
3 - default matrix element is used but one stable particle is of spin$>$0, \\
4 - default matrix element is used with user hadronic current function, \\
5 - user-provided function is used for matrix element calculation. \\
\\
Note that setter for this parameter can only set it to values 0,1,2 or 3.
Values 4 and 5 are reserved for the class constructors and are set based
on the function pointer provided to the constructor.

\item{\tt ChannelForTauola::print()} \\
Prints information about this channel.

\end{description}

\section{Installation and examples}\label{app:install}

The installation procedure of the {\tt tauola-bbb} is similar
to the one of original {\tt TAUOLA}. All relevant source code is located in
{\tt TAUOLA-FORTRAN/tauola-bbb} directory. Executing {\tt make}
in this directory builds all relevant sub-modules and produces the {\tt glib.a}
library%
\footnote{Make sure that {\tt HEPEVT} definition located in
{\tt TAUOLA-FORTRAN/include/HEPEVT.h} is exactly the same as used in
target environment.}.
Executing {\tt make} in any demo subdirectory builds corresponding example.
Running {\tt ./go} script in {\tt prod} subdirectory of any demo directory
runs the example with the default setup provided with the example.

There are three demo subdirectories prepared:
\begin{description}
\item{\tt demo-babar}
\item{\tt demo-lfv}
\item{\tt demo-redefine}
\end{description}

These examples are basically copies of each other with minor differences only.
We decided to split examples into three folders, for user convenience to
grasp properties of new version.

We also provide folders:
\begin{description}
\item{\tt patch-tauolapp}
\item{\tt patch-KK-face}
\item{\tt patch-babar-validation}
\end{description}

These Patch-folders include necessary information on how to use new tarball 
in other projects. The last one is with instruction  how to reproduce 
results of validation for the  new initialization.

In the following Subsections we describe briefly the content of the directories.

\subsection{Default BaBar initialization example}

Directory {\tt demo-babar} is to demonstrate the default initialization of {\tt tauola-bbb}.
This example, produces results validated against BaBar {\tt KKMC}
with BaBar {\tt TAUOLA}. It is the simplest example provided with the
distribution. It does not use {\tt C++} interface for adding channels. 
This example is to show how the default {\tt TAUOLA} output looks like.

\subsection{BaBar initialization used for validation against BaBar data}\label{subsec:valid}

Directory {\tt patch-babar-validation} contains {\tt README} with a list
of modifications that have to be applied to the {\tt tauola-bbb}
in order to allow comparison against BaBar data.

File {\tt babar.root} contains an {\tt MC-TESTER} analysis of the 
 1600~MEvents sample taken from the  production files of BaBar experiment.
File {\tt tauola-bbb.root} contains an analysis of the sample of the same size,
generated using {\tt tauola-bbb} with the modifications described in this
directory. Comparison output {\tt babar.vs.tauola-bbb.pdf} is the result
of the {\tt MC-TESTER} comparison of these two samples.

The instructions located in this directory, show how to reproduce generation
of the {\tt tauola-bbb.root} file. See {\tt MC-TESTER} documentation on
how to perform comparison against {\tt babar.root} sample to produce the {\tt pdf} output.

\subsection{Example of adding LFV current to {\tt tauola-bbb}}

Directory {\tt demo-lfv} expands the {\tt demo-babar} example with
the current described in Section~\ref{sec:Exampleinst}, {\tt README}
located in this directory, describes modifications needed to add this new
current
to the project. In the directory, generation is set for
sample of this single decay channel. 

\subsection{Example of adding new channels and channel redefinition}

Directory {\tt demo-redefine} demonstrates options for the interface
implementation described in Appendix~\ref{app:interface_example}.
File {\tt iniofc.c} located in this directory contains extensive
description of changes that can be introduced using this interface.
This example does not introduce anything valuable in terms of physics
content. Its use is to present the technical side of the project.

\subsubsection{Example of $\tau \to \pi \pi \nu_\tau$ current/matrix element replacement}

The most useful example for learning how the new interface can be used
is to replace one {\tt FORTRAN} channel with it's {\tt C++} version.
In directory {\tt demo-redefine} files {\tt pipi0.c, pipi0.h, MEutils.c MEutils.h}
contain {\tt C++} version of $\tau \to \pi \pi \nu_\tau$ channel and its current.
It is a copy of {\tt FORTRAN} code%
\footnote{We tried to make it as similar to {\tt FORTRAN} counterpart as possible,
so user could compare it step by step.
Note that there are some solutions that do not work in {\tt C++} as a straightforward copies, e.g.
filling 2-dim tables through simple equation, had to be replaced by two {\tt for} loops.}.

User can replace {\tt FORTRAN} hadronic current or the whole matrix element with the {\tt C++} version.
We have performed test to check that the {\tt C++} current gives exactly the same results as its {\tt FORTRAN} counterpart.
With this, user explores baseline to edit the currents,
as well as insight on how hadronic current should look like,
 if they are written from scratch.
We believe, that this example should prepare user for developing his own channels.
To turn this example on or off one should inspect {\tt demo-redefine/iniofc.c}
file.

\subsection{Using {\tt tauola-bbb} with {\tt Tauola Universal Interface}}

Directory {\tt patch-tauolapp} contains {\tt README} file for
the procedure needed to import {\tt tauola-bbb} into the
{\tt C++} Interface of Tauola~\cite{Davidson:2010rw}.
The steps are fairly simple. They boil down to copying the {\tt tauola-bbb}
subdirectory into {\tt TAUOLA/tauola-fortran} subdirectory of the {\tt C++} interface
and applying small patches that allow the directory to be used from the {\tt C++} interface.

\subsection{Using {\tt tauola-bbb} with {\tt KKMC}}

Similarily, directory {\tt patch-KK-face} includes {\tt README} file for
the procedure needed to use {\tt tauola-bbb} with {\tt KKMC}~\cite{Jadach:1999vf}.
The key element is to correctly pass the branching ratio for the new channels
provided by {\tt tauola-bbb}. To achieve this, modifications
described in {\tt README} have to be applied to the default {\tt KKMC} setup. 

Let us provide as an example installation of FCC project~\cite{w1:kkmc,w1:tauolafcc}.
This version is of importance because of plans for the Future Circular Colider of CERN. 
In principle {\tt tauola-bbb/patch-KK-face/README} contains all of the required information, 
however complication due to evolution of software environments have to be taken into account. 
In particular new {\tt Makefile} system of {\tt KKMC} 
requires use of command {\tt alias kmake='make -f KKMakefile'},
which changes use of all old command from {\tt make command} to {\tt kmake command}.

\subsection{New tar-ball modifications}

The distribution tarball discussed in this paper is almost exactly the same
as the version announced in \cite{Was:2014zma}.
Only minor changes were introduced, mostly affecting documentation
(README files and examples of program use) but not the actual code.

In particular:
\begin{enumerate}
\item
{\tt README-changelog} in main {\tt tauola-bbb} directory has been updated
and one incorrect printout in {\tt tauola-c/ChannelForTauola.h} has been fixed.

\item
In file {\tt pkorb.f}, lines 47-48 were added to initialize {\tt BRA1} and {\tt BRKS} to zero.
It was to avoid undefined variables, which for some compilers may not be set to zero anyway, and resolve potential problem with 
CLEO parametrization. Note that this parametrization is presently defunct.

\item
Instead of single {\tt demo-standalone} of older versions of {\tt TAUOLA},
several new, but similar  examples described in previous Subsection have been prepared.

\item
We have 
renamed folders to make clear distinctions between run examples  (demo-*), 
and instructions for patching and validations (patch-*). 
{\tt README} files were updated and some comments were added to account for the
above changes.

\item
Prints for outputs have been updated: 
new version number and release date has been provided.

\item
Minor bugfixing, e.g. prints from {\tt makefile}'s, {\tt KK\_defaults\_adendum.txt}.

\end{enumerate}

\section{Technical prints of default initialization}\label{app:prints}
In previous Appendix an example of how to run the demo programs provided
with the distribution was given.
For the sake of debugging, a function {\tt Tauolapp::PrintChannelInfo(int i)}
have been provided it is declared in {\tt tauola-c/ChannelForTauolaInterface.h} file.
It prints the information, related to the decay channel {\tt i}, and stored 
in the {\tt structs} described in Section~\ref{sec:data}.
One can use this function to monitor the initialization.

The first column of the print,
denotes consecutive channel number, the next its name
which will appear on {\tt TAUOLA} printouts. 
The  {\tt !nu\_tau} denotes that $\nu_\tau$ is
absent from $\tau$ decay products. 
The {\tt b.ratio} denotes probability for the
given channel to be generated. The sum of {\tt b.ratio} over all channels, 
does not need to be equal 1. It is normalized
before event generation anyway. The relative position on the sub-list of
decay channels for the particular phase-space generator follows. Finally
type of the matrix element as explained in Section~\ref{sec:data}, and identifiers
of final state $\tau$ decay products are given. Zero denotes that particular
identifier is not used. This is usually the case of all entries after vertical line $\big|$.
An exception, necessary for anomalous decays,  are position 3 and 4 which are
identifiers for particles to replace default $\nu_\tau$ among $\tau$ decay
products, valid if they follow $\big|$ in the printout of course.

Whenever a pointer to user function for matrix element or current calculation was
properly set, a line:
{\tt      Pointer to user function set correctly.}

\subsection{Default BaBar initialization}

The  printout of the default initialization of BaBar currents,
no user modifications introduced:

{\tiny
\begin{verbatim}
Channel   1: 'TAU-  --> 2PI-,  PI+,  PI0   ' b.ratio: 0.04365 multipl-1: 4 subch:  1 ME-type: 2 F.state:   -1  -1     1     2| 0 0 0 0 0
Channel   2: 'TAU-  --> 3PI0,        PI-   ' b.ratio: 0.01262 multipl-1: 4 subch:  2 ME-type: 2 F.state:    2   2     2    -1| 0 0 0 0 0
Channel   3: 'TAU-  --> nu_e e- e- e+      ' b.ratio: 0.00000 multipl-1: 4 subch:  3 ME-type: 1 F.state:   12 -11   -11    11| 0 0 0 0 0
Channel   4: 'TAU-  --> nu_mu mu- mu- mu+  ' b.ratio: 0.00000 multipl-1: 4 subch:  4 ME-type: 1 F.state:   14 -13   -13    13| 0 0 0 0 0
Channel   5: 'TAU-  --> nu_e e- mu- mu+    ' b.ratio: 0.00000 multipl-1: 4 subch:  5 ME-type: 1 F.state:   12 -11   -13    13| 0 0 0 0 0
Channel   6: 'TAU-  --> nu_mu mu- e- e+    ' b.ratio: 0.00000 multipl-1: 4 subch:  6 ME-type: 1 F.state:   14 -13   -11    11| 0 0 0 0 0
Channel   7: 'TAU-  --> K- 3PI0            ' b.ratio: 0.00000 multipl-1: 4 subch:  7 ME-type: 1 F.state:   -3   2     2     2| 0 0 0 0 0
Channel   8: 'TAU-  --> 2PI0 ETA K-        ' b.ratio: 0.00000 multipl-1: 4 subch:  8 ME-type: 1 F.state:    2   2     9    -3| 0 0 0 0 0
Channel   9: 'TAU-  --> 2PI0 K0  PI-       ' b.ratio: 0.00000 multipl-1: 4 subch:  9 ME-type: 1 F.state:    2   2     4    -1| 0 0 0 0 0
Channel  10: 'TAU-  --> PI0  K0  ETA PI-   ' b.ratio: 0.00000 multipl-1: 4 subch: 10 ME-type: 1 F.state:    2   4     9    -1| 0 0 0 0 0
Channel  11: 'TAU-  --> PI0  PI- PI+ K-    ' b.ratio: 0.00000 multipl-1: 4 subch: 11 ME-type: 1 F.state:    2  -1     1    -3| 0 0 0 0 0
Channel  12: 'TAU-  --> K0   PI- PI+ PI-   ' b.ratio: 0.00000 multipl-1: 4 subch: 12 ME-type: 1 F.state:    4  -1     1    -1| 0 0 0 0 0
Channel  13: 'TAU-  --> 2PI0 ETA PI-       ' b.ratio: 0.00000 multipl-1: 4 subch: 13 ME-type: 1 F.state:    2   2     9    -1| 0 0 0 0 0
Channel  14: 'TAU-  --> K0 K0B ETA PI-     ' b.ratio: 0.00000 multipl-1: 4 subch: 14 ME-type: 1 F.state:    4  -4     9    -1| 0 0 0 0 0
Channel  15: 'TAU-  --> K0 K0B PI0 PI-     ' b.ratio: 0.00000 multipl-1: 4 subch: 15 ME-type: 1 F.state:    4  -4     2    -1| 0 0 0 0 0
Channel  16: 'TAU-  --> K0 K0B K0  PI-     ' b.ratio: 0.00000 multipl-1: 4 subch: 16 ME-type: 1 F.state:    4  -4     4    -1| 0 0 0 0 0
Channel  17: 'TAU-  --> K0 PI0 PI0 K-      ' b.ratio: 0.00000 multipl-1: 4 subch: 17 ME-type: 1 F.state:    4   2     2    -3| 0 0 0 0 0
Channel  18: 'TAU-  --> K0 K0B PI0 K-      ' b.ratio: 0.00000 multipl-1: 4 subch: 18 ME-type: 1 F.state:    4  -4     2    -3| 0 0 0 0 0
Channel  19: 'TAU-  --> PI0  K0  ETA K-    ' b.ratio: 0.00000 multipl-1: 4 subch: 19 ME-type: 1 F.state:    2   4     9    -3| 0 0 0 0 0
Channel  20: 'TAU-  --> PI-PI+PI-  ETA     ' b.ratio: 0.00000 multipl-1: 4 subch: 20 ME-type: 1 F.state:   -1   1    -1     9| 0 0 0 0 0
Channel  21: 'TAU-  --> PI-K+ K-   PI0     ' b.ratio: 0.00000 multipl-1: 4 subch: 21 ME-type: 1 F.state:   -1   3    -3     2| 0 0 0 0 0
Channel  22: 'TAU-  --> K- K+ K-   PI0     ' b.ratio: 0.00000 multipl-1: 4 subch: 22 ME-type: 1 F.state:   -3   3    -3     2| 0 0 0 0 0
Channel  23: 'TAU-  --> K- K+ K-   K0      ' b.ratio: 0.00000 multipl-1: 4 subch: 23 ME-type: 1 F.state:   -3   3    -3     4| 0 0 0 0 0
Channel  24: 'TAU-  --> K- PI+PI-  K0      ' b.ratio: 0.00000 multipl-1: 4 subch: 24 ME-type: 1 F.state:   -3   1    -1     4| 0 0 0 0 0
Channel  25: 'TAU-  --> K- K+ PI-  K0      ' b.ratio: 0.00000 multipl-1: 4 subch: 25 ME-type: 1 F.state:   -3   3    -1     4| 0 0 0 0 0
Channel  26: 'TAU-  --> PI-PI+PI-  OMEGA   ' b.ratio: 0.00000 multipl-1: 4 subch: 26 ME-type: 1 F.state:   -1   1    -1  -223| 0 0 0 0 0
Channel  27: 'TAU-  --> xxxxxxx4xxxxxxxx   ' b.ratio: 0.00000 multipl-1: 4 subch: 27 ME-type: 0 F.state:    4   2     2    -1| 0 0 0 0 0
Channel  28: 'TAU-  --> xxxxxxx4xxxxxxxx   ' b.ratio: 0.00000 multipl-1: 4 subch: 28 ME-type: 0 F.state:    4   2     2    -1| 0 0 0 0 0
Channel  29: 'TAU-  --> xxxxxxx4xxxxxxxx   ' b.ratio: 0.00000 multipl-1: 4 subch: 29 ME-type: 0 F.state:    4   2     2    -1| 0 0 0 0 0
Channel  30: 'TAU-  --> xxxxxxx4xxxxxxxx   ' b.ratio: 0.00000 multipl-1: 4 subch: 30 ME-type: 0 F.state:    4   2     2    -1| 0 0 0 0 0
Channel  31: 'TAU-  --> xxxxxxx4xxxxxxxx   ' b.ratio: 0.00000 multipl-1: 4 subch: 31 ME-type: 0 F.state:    4   2     2    -1| 0 0 0 0 0
Channel  32: 'TAU-  --> xxxxxxx4xxxxxxxx   ' b.ratio: 0.00000 multipl-1: 4 subch: 32 ME-type: 0 F.state:    4   2     2    -1| 0 0 0 0 0
Channel  33: 'TAU-  --> 2PI-, PI+, 2PI0 old' b.ratio: 0.00501 multipl-1: 5 subch:  1 ME-type: 2 F.state:   -1  -1     1     2  2|0 0 0 0
Channel  34: 'TAU-  --> a1 --> rho omega   ' b.ratio: 0.00000 multipl-1: 5 subch:  2 ME-type: 2 F.state:   -1  -1     1     2  2|0 0 0 0
Channel  35: 'TAU-  --> benchmark curr     ' b.ratio: 0.00000 multipl-1: 5 subch:  3 ME-type: 2 F.state:    2   2     2     2  2|0 0 0 0
Channel  36: 'TAU-  --> 2PI- PI+ 2PI0 app08' b.ratio: 0.00000 multipl-1: 5 subch:  4 ME-type: 2 F.state:    1  -1    -1     2  2|0 0 0 0
Channel  37: 'TAU-  --> PI- 4PI0  app08    ' b.ratio: 0.00000 multipl-1: 5 subch:  5 ME-type: 2 F.state:   -1   2     2     2  2|0 0 0 0
Channel  38: 'TAU-  --> 3PI- 2PI+ app08    ' b.ratio: 0.00000 multipl-1: 5 subch:  6 ME-type: 2 F.state:   -1   1     1    -1 -1|0 0 0 0
Channel  39: 'TAU-  --> 2PI- 2PI+  K-      ' b.ratio: 0.00000 multipl-1: 5 subch:  7 ME-type: 1 F.state:   -1  -1     1     1 -3|0 0 0 0
Channel  40: 'TAU-  --> 2PI- PI+ PI0 K0    ' b.ratio: 0.00000 multipl-1: 5 subch:  8 ME-type: 1 F.state:   -1  -1     1     2  4|0 0 0 0
Channel  41: 'TAU-  --> PI- 4PI0           ' b.ratio: 0.00000 multipl-1: 5 subch:  9 ME-type: 2 F.state:   -1   2     2     2  2|0 0 0 0
Channel  42: 'TAU-  --> xxxxxxxxx5xxxxxx   ' b.ratio: 0.00000 multipl-1: 5 subch: 10 ME-type: 0 F.state:   -1  -1     1     2  4|0 0 0 0
Channel  43: 'TAU-  --> xxxxxxxxx5xxxxxx   ' b.ratio: 0.00000 multipl-1: 5 subch: 11 ME-type: 0 F.state:   -1  -1     1     2  4|0 0 0 0
Channel  44: 'TAU-  --> xxxxxxxxx5xxxxxx   ' b.ratio: 0.00000 multipl-1: 5 subch: 12 ME-type: 0 F.state:   -1  -1     1     2  4|0 0 0 0
Channel  45: 'TAU-  --> xxxxxxxxx5xxxxxx   ' b.ratio: 0.00000 multipl-1: 5 subch: 13 ME-type: 0 F.state:   -1  -1     1     2  4|0 0 0 0
Channel  46: 'TAU-  --> xxxxxxxxx5xxxxxx   ' b.ratio: 0.00000 multipl-1: 5 subch: 14 ME-type: 0 F.state:   -1  -1     1     2  4|0 0 0 0
Channel  47: 'TAU-  --> xxxxxxxxx5xxxxxx   ' b.ratio: 0.00000 multipl-1: 5 subch: 15 ME-type: 0 F.state:   -1  -1     1     2  4|0 0 0 0
Channel  48: 'TAU-  --> xxxxxxxxx5xxxxxx   ' b.ratio: 0.00000 multipl-1: 5 subch: 16 ME-type: 0 F.state:   -1  -1     1     2  4|0 0 0 0
Channel  49: 'TAU-  --> xxxxxxxxx5xxxxxx   ' b.ratio: 0.00000 multipl-1: 5 subch: 17 ME-type: 0 F.state:   -1  -1     1     2  4|0 0 0 0
Channel  50: 'TAU-  --> xxxxxxxxx5xxxxxx   ' b.ratio: 0.00000 multipl-1: 5 subch: 18 ME-type: 0 F.state:   -1  -1     1     2  4|0 0 0 0
Channel  51: 'TAU-  --> xxxxxxxxx5xxxxxx   ' b.ratio: 0.00000 multipl-1: 5 subch: 19 ME-type: 0 F.state:   -1  -1     1     2  4|0 0 0 0
Channel  52: 'TAU-  --> xxxxxxxxx5xxxxxx   ' b.ratio: 0.00000 multipl-1: 5 subch: 20 ME-type: 0 F.state:   -1  -1     1     2  4|0 0 0 0
Channel  53: 'TAU-  --> xxxxxxxxx5xxxxxx   ' b.ratio: 0.00000 multipl-1: 5 subch: 21 ME-type: 0 F.state:   -1  -1     1     2  4|0 0 0 0
Channel  54: 'TAU-  --> 3PI-, 2PI+,        ' b.ratio: 0.00079 multipl-1: 5 subch:  1 ME-type: 2 F.state:   -1  -1    -1     1  1|0 0 0 0
Channel  55: 'TAU-  --> 3PI-, 2PI+,  PI0   ' b.ratio: 0.00018 multipl-1: 6 subch:  2 ME-type: 2 F.state:   -1  -1    -1     1  1 2|0 0 0
Channel  56: 'TAU-  --> 2PI-,  PI+, 3PI0   ' b.ratio: 0.00025 multipl-1: 6 subch:  3 ME-type: 2 F.state:   -1  -1     1     2  2 2|0 0 0
Channel  57: 'TAU-  --> 3pi- 2pi+ 2pi0     ' b.ratio: 0.00000 multipl-1: 7 subch:  4 ME-type: 2 F.state:   -1  -1    -1     1  1 2 2|0 0
Channel  58: 'TAU-  --> 4PI- 3PI+          ' b.ratio: 0.00000 multipl-1: 7 subch:  5 ME-type: 2 F.state:   -1  -1    -1    -1  1 1 1|0 0
Channel  59: 'TAU-  --> 4PI- 3PI+  PI0     ' b.ratio: 0.00000 multipl-1: 8 subch:  6 ME-type: 2 F.state:   -1  -1    -1    -1  1 1 1 2|0
Channel  60: 'TAU-  --> 2PI- 2PI+ K- PI0   ' b.ratio: 0.00000 multipl-1: 6 subch:  7 ME-type: 2 F.state:   -1  -1     1     1 -3 2|0 0 0
Channel  61: 'TAU-  --> xxxxxxxxxnxxxxxx   ' b.ratio: 0.00000 multipl-1: 6 subch:  8 ME-type: 0 F.state:   -1  -1    -1     1  1 1|0 0 0
Channel  62: 'TAU-  --> xxxxxxxxxnxxxxxx   ' b.ratio: 0.00000 multipl-1: 6 subch:  9 ME-type: 0 F.state:   -1  -1     1     2  2 1|0 0 0
Channel  63: 'TAU-  --> xxxxxxxxxnxxxxxx   ' b.ratio: 0.00000 multipl-1: 6 subch: 10 ME-type: 0 F.state:   -1  -1    -1     1  1 1|0 0 0
Channel  64: 'TAU-  --> xxxxxxxxxnxxxxxx   ' b.ratio: 0.00000 multipl-1: 6 subch: 11 ME-type: 0 F.state:   -1  -1     1     2  2 1|0 0 0
Channel  65: 'TAU-  --> xxxxxxxxxnxxxxxx   ' b.ratio: 0.00000 multipl-1: 6 subch: 12 ME-type: 0 F.state:   -1  -1    -1     1  1 1|0 0 0
Channel  66: 'TAU-  --> xxxxxxxxxnxxxxxx   ' b.ratio: 0.00000 multipl-1: 6 subch: 13 ME-type: 0 F.state:   -1  -1     1     2  2 1|0 0 0
Channel  67: 'TAU-  -->  K-, PI-,  K+      ' b.ratio: 0.00159 multipl-1: 3 subch:  1 ME-type: 2 F.state:   -3  -1     3|    0  0 0 0 0 0
Channel  68: 'TAU-  -->  K0, PI-, K0B      ' b.ratio: 0.00167 multipl-1: 3 subch:  2 ME-type: 2 F.state:   -4  -1     4|    0  0 0 0 0 0
Channel  69: 'TAU-  -->  K-,  PI0, K0      ' b.ratio: 0.00154 multipl-1: 3 subch:  3 ME-type: 2 F.state:   -3   2    -4|    0  0 0 0 0 0
Channel  70: 'TAU-  --> PI0  PI0   K-      ' b.ratio: 0.00068 multipl-1: 3 subch:  4 ME-type: 2 F.state:    2   2    -3|    0  0 0 0 0 0
Channel  71: 'TAU-  -->  K-  PI-  PI+      ' b.ratio: 0.00301 multipl-1: 3 subch:  5 ME-type: 2 F.state:   -3  -1     1|    0  0 0 0 0 0
Channel  72: 'TAU-  --> PI-  K0B  PI0      ' b.ratio: 0.00377 multipl-1: 3 subch:  6 ME-type: 2 F.state:   -1   4     2|    0  0 0 0 0 0
Channel  73: 'TAU-  --> ETA  PI-  PI0      ' b.ratio: 0.00183 multipl-1: 3 subch:  7 ME-type: 2 F.state:    9  -1     2|    0  0 0 0 0 0
Channel  74: 'TAU-  --> PI-  PI0  GAM      ' b.ratio: 0.00080 multipl-1: 3 subch:  8 ME-type: 3 F.state:   -1   2     8|    0  0 0 0 0 0
Channel  75: 'TAU-  --> PI0  PI0  PI-      ' b.ratio: 0.09178 multipl-1: 3 subch:  9 ME-type: 2 F.state:    2   2    -1|    0  0 0 0 0 0
Channel  76: 'TAU-  --> PI-  PI-  PI+      ' b.ratio: 0.09178 multipl-1: 3 subch: 10 ME-type: 2 F.state:   -1  -1     1|    0  0 0 0 0 0
Channel  77: 'TAU-  --> K-    K-   K+      ' b.ratio: 0.00000 multipl-1: 3 subch: 11 ME-type: 1 F.state:   -3  -3     3|    0  0 0 0 0 0
Channel  78: 'TAU-  --> K-    K0   K0      ' b.ratio: 0.00000 multipl-1: 3 subch: 12 ME-type: 1 F.state:   -3   4     4|    0  0 0 0 0 0
Channel  79: 'TAU-  --> K-   ETA  PI0      ' b.ratio: 0.00000 multipl-1: 3 subch: 13 ME-type: 1 F.state:   -3   9     2|    0  0 0 0 0 0
Channel  80: 'TAU-  --> K0   ETA  PI-      ' b.ratio: 0.00000 multipl-1: 3 subch: 14 ME-type: 1 F.state:    4   9    -1|    0  0 0 0 0 0
Channel  81: 'TAU-  --> K-   K0   RHO0     ' b.ratio: 0.00000 multipl-1: 3 subch: 15 ME-type: 1 F.state:   -3   4   113|    0  0 0 0 0 0
Channel  82: 'TAU-  --> PI-  PHI  PI0      ' b.ratio: 0.00000 multipl-1: 3 subch: 16 ME-type: 1 F.state:   -1 333     2|    0  0 0 0 0 0
Channel  83: 'TAU-  --> K-   PHI  PI0      ' b.ratio: 0.00000 multipl-1: 3 subch: 17 ME-type: 1 F.state:   -3 333     2|    0  0 0 0 0 0
Channel  84: 'TAU-  --> K0   ETA  K-       ' b.ratio: 0.00000 multipl-1: 3 subch: 18 ME-type: 1 F.state:    4   9    -3|    0  0 0 0 0 0
Channel  85: 'TAU-  --> xxxxxxxxx3xxxxxx   ' b.ratio: 0.00000 multipl-1: 3 subch: 19 ME-type: 0 F.state:    2   2     2|    0  0 0 0 0 0
Channel  86: 'TAU-  -->  PI- PI0           ' b.ratio: 0.25375 multipl-1: 2 subch:  1 ME-type: 2 F.state:   -1   2|    0     0  0 0 0 0 0
Channel  87: 'TAU-  -->  PI- K0            ' b.ratio: 0.00909 multipl-1: 2 subch:  2 ME-type: 2 F.state:   -1  -4|    0     0  0 0 0 0 0
Channel  88: 'TAU-  -->  K-  PI0           ' b.ratio: 0.00455 multipl-1: 2 subch:  3 ME-type: 2 F.state:   -3   2|    0     0  0 0 0 0 0
Channel  89: 'TAU-  -->  K-  K0            ' b.ratio: 0.00165 multipl-1: 2 subch:  4 ME-type: 2 F.state:   -3  -4|    0     0  0 0 0 0 0
Channel  90: 'TAU-  -->  mu-mu-mu+ !nu_tau ' b.ratio: 0.00000 multipl-1: 2 subch:  5 ME-type: 1 F.state:  -13 -13|   13     0  0 0 0 0 0
Channel  91: 'TAU-  --> mu- mu- e+ !nu_tau ' b.ratio: 0.00000 multipl-1: 2 subch:  6 ME-type: 1 F.state:  -13 -13|   11     0  0 0 0 0 0
Channel  92: 'TAU-  --> mu- e- mu+ !nu_tau ' b.ratio: 0.00000 multipl-1: 2 subch:  7 ME-type: 1 F.state:  -13 -11|   13     0  0 0 0 0 0
Channel  93: 'TAU-  --> mu- e- e+  !nu_tau ' b.ratio: 0.00000 multipl-1: 2 subch:  8 ME-type: 1 F.state:  -13 -11|   11     0  0 0 0 0 0
Channel  94: 'TAU-  --> mu+ e- e-  !nu_tau ' b.ratio: 0.00000 multipl-1: 2 subch:  9 ME-type: 1 F.state:   13 -11|   11     0  0 0 0 0 0
Channel  95: 'TAU-  --> e- e- e+   !nu_tau ' b.ratio: 0.00000 multipl-1: 2 subch: 10 ME-type: 1 F.state:  -11 -11|   11     0  0 0 0 0 0
Channel  96: 'TAU-  --> e-pi+pi-  !nu_tau  ' b.ratio: 0.00000 multipl-1: 2 subch: 11 ME-type: 1 F.state:   -1   1|  -11     0  0 0 0 0 0
Channel  97: 'TAU-  --> mu-pi+pi-  !nu_tau ' b.ratio: 0.00000 multipl-1: 2 subch: 12 ME-type: 1 F.state:   -1   1|  -13     0  0 0 0 0 0
Channel  98: 'TAU-  --> e-pi+K-  !nu_tau   ' b.ratio: 0.00000 multipl-1: 2 subch: 13 ME-type: 1 F.state:    1  -3|  -11     0  0 0 0 0 0
Channel  99: 'TAU-  --> mu-pi+K-  !nu_tau  ' b.ratio: 0.00000 multipl-1: 2 subch: 14 ME-type: 1 F.state:    1  -3|  -13     0  0 0 0 0 0
Channel 100: 'TAU-  --> e-pi-K+  !nu_tau   ' b.ratio: 0.00000 multipl-1: 2 subch: 15 ME-type: 1 F.state:   -1   3|  -11     0  0 0 0 0 0
Channel 101: 'TAU-  --> mu-pi-K+  !nu_tau  ' b.ratio: 0.00000 multipl-1: 2 subch: 16 ME-type: 1 F.state:   -1   3|  -13     0  0 0 0 0 0
Channel 102: 'TAU-  --> e-K-K+  !nu_tau    ' b.ratio: 0.00000 multipl-1: 2 subch: 17 ME-type: 1 F.state:   -3   3|  -11     0  0 0 0 0 0
Channel 103: 'TAU-  --> mu-K-K+  !nu_tau   ' b.ratio: 0.00000 multipl-1: 2 subch: 18 ME-type: 1 F.state:   -3   3|  -13     0  0 0 0 0 0
Channel 104: 'TAU-  --> e-K0K0  !nu_tau    ' b.ratio: 0.00000 multipl-1: 2 subch: 19 ME-type: 1 F.state:    4   4|  -11     0  0 0 0 0 0
Channel 105: 'TAU-  --> mu-K0K0  !nu_tau   ' b.ratio: 0.00000 multipl-1: 2 subch: 20 ME-type: 1 F.state:    4   4|  -13     0  0 0 0 0 0
Channel 106: 'TAU-  --> e+pi-pi-  !nu_tau  ' b.ratio: 0.00000 multipl-1: 2 subch: 21 ME-type: 1 F.state:   -1  -1|   11     0  0 0 0 0 0
Channel 107: 'TAU-  --> mu+pi-pi-  !nu_tau ' b.ratio: 0.00000 multipl-1: 2 subch: 22 ME-type: 1 F.state:   -1  -1|   13     0  0 0 0 0 0
Channel 108: 'TAU-  --> e+pi-K-  !nu_tau   ' b.ratio: 0.00000 multipl-1: 2 subch: 23 ME-type: 1 F.state:   -1  -3|   11     0  0 0 0 0 0
Channel 109: 'TAU-  --> mu+pi-K-  !nu_tau  ' b.ratio: 0.00000 multipl-1: 2 subch: 24 ME-type: 1 F.state:   -1  -3|   13     0  0 0 0 0 0
Channel 110: 'TAU-  --> e+K-K-  !nu_tau    ' b.ratio: 0.00000 multipl-1: 2 subch: 25 ME-type: 1 F.state:   -3  -3|   11     0  0 0 0 0 0
Channel 111: 'TAU-  --> mu+K-K-  !nu_tau   ' b.ratio: 0.00000 multipl-1: 2 subch: 26 ME-type: 1 F.state:   -3  -3|   13     0  0 0 0 0 0
Channel 112: 'TAU-  --> mu-mu- p+  !nu_tau ' b.ratio: 0.00000 multipl-1: 2 subch: 27 ME-type: 1 F.state:  -13 -13|-2212     0  0 0 0 0 0
Channel 113: 'TAU-  --> mu-mu+ p-  !nu_tau ' b.ratio: 0.00000 multipl-1: 2 subch: 28 ME-type: 1 F.state:  -13  13| 2212     0  0 0 0 0 0
Channel 114: 'TAU-  -->  e - e- p+ !nu_tau ' b.ratio: 0.00000 multipl-1: 2 subch: 29 ME-type: 1 F.state:  -11 -11|-2212     0  0 0 0 0 0
Channel 115: 'TAU-  -->  e - e+ p- !nu_tau ' b.ratio: 0.00000 multipl-1: 2 subch: 30 ME-type: 1 F.state:  -11  11| 2212     0  0 0 0 0 0
Channel 116: 'TAU-  --> eta k-             ' b.ratio: 0.00000 multipl-1: 2 subch: 31 ME-type: 1 F.state:    9  -3|    0     0  0 0 0 0 0
Channel 117: 'TAU-  --> eta pi-            ' b.ratio: 0.00000 multipl-1: 2 subch: 32 ME-type: 1 F.state:    9  -1|    0     0  0 0 0 0 0
Channel 118: 'TAU-  --> PI-  PHI           ' b.ratio: 0.00000 multipl-1: 2 subch: 33 ME-type: 1 F.state:   -1 333|    0     0  0 0 0 0 0
Channel 119: 'TAU-  -->  K-  PHI           ' b.ratio: 0.00000 multipl-1: 2 subch: 34 ME-type: 1 F.state:   -3 333|    0     0  0 0 0 0 0
Channel 120: 'TAU-  -->  PI- OMEGA         ' b.ratio: 0.00000 multipl-1: 2 subch: 35 ME-type: 1 F.state:   -1 223|    0     0  0 0 0 0 0
Channel 121: 'TAU-  -->  K-  OMEGA         ' b.ratio: 0.00000 multipl-1: 2 subch: 36 ME-type: 1 F.state:   -3 223|    0     0  0 0 0 0 0
Channel 122: 'TAU-  -->  PI- ETAprm        ' b.ratio: 0.00000 multipl-1: 2 subch: 37 ME-type: 1 F.state:   -1 331|    0     0  0 0 0 0 0
Channel 123: 'TAU-  -->  K-  ETAprm        ' b.ratio: 0.00000 multipl-1: 2 subch: 38 ME-type: 1 F.state:   -3 331|    0     0  0 0 0 0 0
Channel 124: 'TAU-  -->  e- mu+ p- !nu_tau ' b.ratio: 0.00000 multipl-1: 2 subch: 39 ME-type: 1 F.state:  -11  13| 2212     0  0 0 0 0 0
Channel 125: 'TAU-  -->  e+ mu- p- !nu_tau ' b.ratio: 0.00000 multipl-1: 2 subch: 40 ME-type: 1 F.state:   11 -13| 2212     0  0 0 0 0 0
Channel 126: 'TAU-  -->  e- mu- p+ !nu_tau ' b.ratio: 0.00000 multipl-1: 2 subch: 41 ME-type: 1 F.state:  -11 -13|-2212     0  0 0 0 0 0
Channel 127: 'TAU-  --> e- PI0 PI0  !nu_tau' b.ratio: 0.00000 multipl-1: 2 subch: 42 ME-type: 1 F.state:    2   2|  -11     0  0 0 0 0 0
Channel 128: 'TAU-  --> mu- PI0 PI0 !nu_tau' b.ratio: 0.00000 multipl-1: 2 subch: 43 ME-type: 1 F.state:    2   2|  -13     0  0 0 0 0 0
Channel 129: 'TAU-  --> e- PI0 eta !nu_tau ' b.ratio: 0.00000 multipl-1: 2 subch: 44 ME-type: 1 F.state:    2   9|  -11     0  0 0 0 0 0
Channel 130: 'TAU-  --> mu- PI0 eta !nu_tau' b.ratio: 0.00000 multipl-1: 2 subch: 45 ME-type: 1 F.state:    2   9|  -13     0  0 0 0 0 0
Channel 131: 'TAU-  -->  e- PI0 eta_p !nu_t' b.ratio: 0.00000 multipl-1: 2 subch: 46 ME-type: 1 F.state:    2 331|  -11     0  0 0 0 0 0
Channel 132: 'TAU-  --> mu- PI0 eta_p !nu_t' b.ratio: 0.00000 multipl-1: 2 subch: 47 ME-type: 1 F.state:    2 331|  -13     0  0 0 0 0 0
Channel 133: 'TAU-  --> e- eta eta  !nu_tau' b.ratio: 0.00000 multipl-1: 2 subch: 48 ME-type: 1 F.state:    9   9|  -11     0  0 0 0 0 0
Channel 134: 'TAU-  --> mu- eta eta !nu_tau' b.ratio: 0.00000 multipl-1: 2 subch: 49 ME-type: 1 F.state:    9   9|  -13     0  0 0 0 0 0
Channel 135: 'TAU-  --> e- eta eta_p !nu_t ' b.ratio: 0.00000 multipl-1: 2 subch: 50 ME-type: 1 F.state:    9 331|  -11     0  0 0 0 0 0
Channel 136: 'TAU-  --> mu- eta eta_p !nu_t' b.ratio: 0.00000 multipl-1: 2 subch: 51 ME-type: 1 F.state:    9 331|  -13     0  0 0 0 0 0
Channel 137: 'TAU-  --> e- PI0 Ks  !nu_tau ' b.ratio: 0.00000 multipl-1: 2 subch: 52 ME-type: 1 F.state:    2 310|  -11     0  0 0 0 0 0
Channel 138: 'TAU-  --> mu- PI0 Ks !nu_tau ' b.ratio: 0.00000 multipl-1: 2 subch: 53 ME-type: 1 F.state:    2 310|  -13     0  0 0 0 0 0
Channel 139: 'TAU-  --> e- eta  Ks !nu_tau ' b.ratio: 0.00000 multipl-1: 2 subch: 54 ME-type: 1 F.state:    9 310|  -11     0  0 0 0 0 0
Channel 140: 'TAU-  --> mu- eta Ks !nu_tau ' b.ratio: 0.00000 multipl-1: 2 subch: 55 ME-type: 1 F.state:    9 310|  -13     0  0 0 0 0 0
Channel 141: 'TAU-  --> e- eta_p Ks !nu_tau' b.ratio: 0.00000 multipl-1: 2 subch: 56 ME-type: 1 F.state:  331 310|  -11     0  0 0 0 0 0
Channel 142: 'TAU-  --> mu- eta_p Ks !nu_t ' b.ratio: 0.00000 multipl-1: 2 subch: 57 ME-type: 1 F.state:  331 310|  -13     0  0 0 0 0 0
Channel 143: 'TAU-  --> p- pi+ K- !nu_tau  ' b.ratio: 0.00000 multipl-1: 2 subch: 58 ME-type: 1 F.state:    1  -3| 2212     0  0 0 0 0 0
Channel 144: 'TAU-  --> p+ pi- K-  !nu_tau ' b.ratio: 0.00000 multipl-1: 2 subch: 59 ME-type: 0 F.state:   -1  -3|-2212     0  0 0 0 0 0
Channel 145: 'TAU-  --> p- K+ pi- !nu_tau  ' b.ratio: 0.00000 multipl-1: 2 subch: 60 ME-type: 0 F.state:    3  -1| 2212     0  0 0 0 0 0
Channel 146: 'TAU-  --> p- pi0 pi0 !nu_tau ' b.ratio: 0.00000 multipl-1: 2 subch: 61 ME-type: 0 F.state:    2   2| 2212     0  0 0 0 0 0
Channel 147: 'TAU-  --> p- pi0 eta !nu_tau ' b.ratio: 0.00000 multipl-1: 2 subch: 62 ME-type: 0 F.state:    2   9| 2212     0  0 0 0 0 0
Channel 148: 'TAU-  --> p- pi0 Ks !nu_tau  ' b.ratio: 0.00000 multipl-1: 2 subch: 63 ME-type: 0 F.state:    2 310| 2212     0  0 0 0 0 0
Channel 149: 'TAU-  --> xxxxxxxxx2xxxxxx   ' b.ratio: 0.00000 multipl-1: 2 subch: 64 ME-type: 0 F.state:   -3  -3|    0     0  0 0 0 0 0
Channel 150: 'TAU-  --> xxxxxxxxx2xxxxxx   ' b.ratio: 0.00000 multipl-1: 2 subch: 65 ME-type: 0 F.state:   -3  -3|    0     0  0 0 0 0 0
Channel 151: 'TAU-  --> xxxxxxxxx2xxxxxx   ' b.ratio: 0.00000 multipl-1: 2 subch: 66 ME-type: 0 F.state:   -3  -3|    0     0  0 0 0 0 0
Channel 152: 'TAU-  --> xxxxxxxxx2xxxxxx   ' b.ratio: 0.00000 multipl-1: 2 subch: 67 ME-type: 0 F.state:   -3  -3|    0     0  0 0 0 0 0
Channel 153: 'TAU-  --> xxxxxxxxx2xxxxxx   ' b.ratio: 0.00000 multipl-1: 2 subch: 68 ME-type: 0 F.state:   -3  -3|    0     0  0 0 0 0 0
Channel 154: 'TAU-  --> xxxxxxxxx2xxxxxx   ' b.ratio: 0.00000 multipl-1: 2 subch: 69 ME-type: 0 F.state:   -3  -3|    0     0  0 0 0 0 0
Channel 155: 'TAU-  --> xxxxxxxxx2xxxxxx   ' b.ratio: 0.00000 multipl-1: 2 subch: 70 ME-type: 0 F.state:   -3  -3|    0     0  0 0 0 0 0
Channel 156: 'TAU-  --> xxxxxxxxx2xxxxxx   ' b.ratio: 0.00000 multipl-1: 2 subch: 71 ME-type: 0 F.state:   -3  -3|    0     0  0 0 0 0 0
Channel 157: 'TAU-  --> PI-                ' b.ratio: 0.11084 multipl-1: 1 subch:  1 ME-type: 2 F.state:   -1|  0     0     0  0 0 0 0 0
Channel 158: 'TAU-  --> K-                 ' b.ratio: 0.00695 multipl-1: 1 subch:  2 ME-type: 2 F.state:   -3|  0     0     0  0 0 0 0 0
Channel 159: 'TAU-  --> gamma e-   !nu_tau ' b.ratio: 0.00000 multipl-1: 1 subch:  3 ME-type: 1 F.state:    8|  0   -11     0  0 0 0 0 0
Channel 160: 'TAU-  --> gamma mu-  !nu_tau ' b.ratio: 0.00000 multipl-1: 1 subch:  4 ME-type: 1 F.state:    8|  0   -13     0  0 0 0 0 0
Channel 161: 'TAU-  --> PI0 e-     !nu_tau ' b.ratio: 0.00000 multipl-1: 1 subch:  5 ME-type: 1 F.state:    2|  0   -11     0  0 0 0 0 0
Channel 162: 'TAU-  --> PI0 mu-    !nu_tau ' b.ratio: 0.00000 multipl-1: 1 subch:  6 ME-type: 1 F.state:    2|  0   -13     0  0 0 0 0 0
Channel 163: 'TAU-  --> eta e-     !nu_tau ' b.ratio: 0.00000 multipl-1: 1 subch:  7 ME-type: 1 F.state:    9|  0   -11     0  0 0 0 0 0
Channel 164: 'TAU-  --> eta mu-    !nu_tau ' b.ratio: 0.00000 multipl-1: 1 subch:  8 ME-type: 1 F.state:    9|  0   -13     0  0 0 0 0 0
Channel 165: 'TAU-  --> e-  K0     !nu_tau ' b.ratio: 0.00000 multipl-1: 1 subch:  9 ME-type: 1 F.state:    4|  0   -11     0  0 0 0 0 0
Channel 166: 'TAU-  --> mu- K0     !nu_tau ' b.ratio: 0.00000 multipl-1: 1 subch: 10 ME-type: 1 F.state:    4|  0   -13     0  0 0 0 0 0
Channel 167: 'TAU-  --> e-  omega  !nu_tau ' b.ratio: 0.00000 multipl-1: 1 subch: 11 ME-type: 1 F.state:  -11|  0     0   223  0 0 0 0 0
Channel 168: 'TAU-  --> mu- omega  !nu_tau ' b.ratio: 0.00000 multipl-1: 1 subch: 12 ME-type: 1 F.state:  -13|  0     0   223  0 0 0 0 0
Channel 169: 'TAU-  --> e-  phi    !nu_tau ' b.ratio: 0.00000 multipl-1: 1 subch: 13 ME-type: 1 F.state:  -11|  0     0   333  0 0 0 0 0
Channel 170: 'TAU-  --> mu- phi    !nu_tau ' b.ratio: 0.00000 multipl-1: 1 subch: 14 ME-type: 1 F.state:  -13|  0     0   333  0 0 0 0 0
Channel 171: 'TAU-  --> e- rho0    !nu_tau ' b.ratio: 0.00000 multipl-1: 1 subch: 15 ME-type: 1 F.state:  -11|  0     0   113  0 0 0 0 0
Channel 172: 'TAU-  --> mu- rho0   !nu_tau ' b.ratio: 0.00000 multipl-1: 1 subch: 16 ME-type: 1 F.state:  -13|  0     0   113  0 0 0 0 0
Channel 173: 'TAU-  --> A0-                ' b.ratio: 0.00000 multipl-1: 1 subch: 17 ME-type: 1 F.state:10211|  0     0     0  0 0 0 0 0
Channel 174: 'TAU-  --> B1-                ' b.ratio: 0.00000 multipl-1: 1 subch: 18 ME-type: 1 F.state:10213|  0     0     0  0 0 0 0 0
Channel 175: 'TAU-  --> e- K0    !nu_tau   ' b.ratio: 0.00000 multipl-1: 1 subch: 19 ME-type: 1 F.state:  -11|  0     0   311  0 0 0 0 0
Channel 176: 'TAU-  --> mu- K0    !nu_tau  ' b.ratio: 0.00000 multipl-1: 1 subch: 20 ME-type: 1 F.state:  -13|  0     0   311  0 0 0 0 0
Channel 177: 'TAU-  -->  p gamma  !nu_tau  ' b.ratio: 0.00000 multipl-1: 1 subch: 21 ME-type: 1 F.state: 2212|  0     0    22  0 0 0 0 0
Channel 178: 'TAU-  --> p pi0     !nu_tau  ' b.ratio: 0.00000 multipl-1: 1 subch: 22 ME-type: 1 F.state: 2212|  0     0   111  0 0 0 0 0
Channel 179: 'TAU-  --> p eta    !nu_tau   ' b.ratio: 0.00000 multipl-1: 1 subch: 23 ME-type: 1 F.state: 2212|  0     0   221  0 0 0 0 0
Channel 180: 'TAU-  -->  p K0   !nu_tau    ' b.ratio: 0.00000 multipl-1: 1 subch: 24 ME-type: 1 F.state: 2212|  0     0   311  0 0 0 0 0
Channel 181: 'TAU-  --> e- eta_p  !nu_tau  ' b.ratio: 0.00000 multipl-1: 1 subch: 25 ME-type: 1 F.state:  -11|  0     0   331  0 0 0 0 0
Channel 182: 'TAU-  --> mu- eta_p !nu_tau  ' b.ratio: 0.00000 multipl-1: 1 subch: 26 ME-type: 1 F.state:  -13|  0     0   331  0 0 0 0 0
Channel 183: 'TAU-  --> pi- lambda !nu_tau ' b.ratio: 0.00000 multipl-1: 1 subch: 27 ME-type: 1 F.state:   -1|  0     0  3122  0 0 0 0 0
Channel 184: 'TAU-  --> pi- lmb_br !nu_tau ' b.ratio: 0.00000 multipl-1: 1 subch: 28 ME-type: 1 F.state:   -1|  0     0 -3122  0 0 0 0 0
Channel 185: 'TAU-  --> K- lambda  !nu_tau ' b.ratio: 0.00000 multipl-1: 1 subch: 29 ME-type: 1 F.state:   -3|  0     0  3122  0 0 0 0 0
Channel 186: 'TAU-  --> K- lmb_bar !nu_tau ' b.ratio: 0.00000 multipl-1: 1 subch: 30 ME-type: 1 F.state:   -3|  0     0 -3122  0 0 0 0 0
Channel 187: 'TAU-  --> e-  K*  !nu_tau    ' b.ratio: 0.00000 multipl-1: 1 subch: 31 ME-type: 1 F.state:  -11|  0     0   313  0 0 0 0 0
Channel 188: 'TAU-  --> e-  K*_bar !nu_tau ' b.ratio: 0.00000 multipl-1: 1 subch: 32 ME-type: 1 F.state:  -11|  0     0  -313  0 0 0 0 0
Channel 189: 'TAU-  --> mu- K*_bar !nu_tau ' b.ratio: 0.00000 multipl-1: 1 subch: 33 ME-type: 1 F.state:  -13|  0     0   313  0 0 0 0 0
Channel 190: 'TAU-  --> mu-  K*  !nu_tau   ' b.ratio: 0.00000 multipl-1: 1 subch: 34 ME-type: 1 F.state:  -13|  0     0  -313  0 0 0 0 0
Channel 191: 'TAU-  --> e- a0(980) !nu_tau ' b.ratio: 0.00000 multipl-1: 1 subch: 35 ME-type: 1 F.state:  -11|  0     0 10111  0 0 0 0 0
Channel 192: 'TAU-  --> mu- a0(980) !nu_tau' b.ratio: 0.00000 multipl-1: 1 subch: 36 ME-type: 1 F.state:  -13|  0     0 10111  0 0 0 0 0
Channel 193: 'TAU-  --> e-  f0(980) !nu_tau' b.ratio: 0.00000 multipl-1: 1 subch: 37 ME-type: 1 F.state:  -11|  0     0 10221  0 0 0 0 0
Channel 194: 'TAU-  --> mu- f0(980) !nu_tau' b.ratio: 0.00000 multipl-1: 1 subch: 38 ME-type: 1 F.state:  -13|  0     0 10221  0 0 0 0 0
Channel 195: 'TAU-  --> xxxxxxxxx1xxxxxx   ' b.ratio: 0.00000 multipl-1: 1 subch: 39 ME-type: 0 F.state:   -3|  0     0     0  0 0 0 0 0
Channel 196: 'TAU-  --> xxxxxxxxx1xxxxxx   ' b.ratio: 0.00000 multipl-1: 1 subch: 40 ME-type: 0 F.state:   -3|  0     0     0  0 0 0 0 0
\end{verbatim}
}

\end{document}